\documentclass[10pt,journal,compsoc]{IEEEtran}
\ifCLASSINFOpdf
  \usepackage[pdftex]{graphicx}
  \graphicspath{{./img/}}
  \DeclareGraphicsExtensions{.pdf,.jpeg,.png}
\else
  \usepackage[dvips]{graphicx}
  \graphicspath{{./img/}}
  \DeclareGraphicsExtensions{.eps}
\fi
\usepackage{forest}

\usepackage[cmex10]{amsmath}

\usepackage[]{algorithm2e}

\ifCLASSOPTIONcompsoc
  \usepackage[caption=false,font=normalsize,labelfont=sf,textfont=sf]{subfig}
\else
  \usepackage[caption=false,font=footnotesize]{subfig}
\fi
\usepackage{url}
\usepackage[normalem]{ulem}

\usepackage[]{comment}
\usepackage[inline]{enumitem}

\newcommand{\CommentBegin}[1] {\textbf{*** BEGIN ***}}
\newcommand{\CommentEnd}[1] {\textbf{*** END ***}}

\usepackage{prettyref}
\newrefformat{fig}{Figure~\ref{#1}}
\newrefformat{tab}{Table~\ref{#1}}
\newrefformat{sec}{Section~\ref{#1}}
\newrefformat{ssec}{Section~\ref{#1}}
\newrefformat{subsec}{Section~\ref{#1}}
\newrefformat{alg}{Algorithm~\ref{#1}}
\newrefformat{eq}{Eq.~\ref{#1}}
\newrefformat{line}{Line~\ref{#1}}

\newcommand{\MiniApp}{\emph{GeoDock-MA}}

\begin{document}

\title{\huge Tunable Approximations to Control Time-to-Solution in \\ an HPC Molecular Docking Mini-App}

\author{
Davide~Gadioli,
Gianluca~Palermo,
Stefano~Cherubin,
Emanuele~Vitali,
Giovanni~Agosta,
Candida~Manelfi,
Andrea~R.~Beccari,
Carlo~Cavazzoni,
Nico~Sanna,
and Cristina~Silvano
\IEEEcompsocitemizethanks{
\IEEEcompsocthanksitem D. Gadioli, G. Palermo, S. Cherubin, E. Vitali, G. Agosta and C. Silvano are with Dipartimento di Elettronica, Infomazione e Bioingegneria,
Politecnico di Milano, Italy.
\IEEEcompsocthanksitem C. Manelfi and A. Beccari are with Dompe' Farmaceutici SpA Research Center, L'Aquila, Italy.
\IEEEcompsocthanksitem C. Cavazzoni and N. Sanna are with the Supercomputing Innovation and Application Department, CINECA, Bologna/Roma, Italy.
}

\thanks{This work is supported by the European Union's Horizon 2020 research and innovation program under grant agreement No 671623, FET-HPC ANTAREX.}
}


\IEEEtitleabstractindextext{%
\begin{abstract}

The drug discovery process involves several tasks to be performed \textit{in vivo}, \textit{in vitro} and \textit{in silico}.
Molecular docking is a task typically performed \textit{in silico}.
It aims at finding the three-dimensional pose of a given molecule when it interacts with the target protein binding site.
This task is often done for virtual screening a huge set of molecules to find the most promising ones, which will be forwarded to the later stages of the drug discovery process.
Given the huge complexity of the problem, molecular docking cannot be solved by exploring the entire space of the ligand poses.
State-of-the-art approaches face the problem by sampling the space of the ligand poses to generate results in a reasonable time budget.
In this work, we improve the geometric approach to molecular docking by introducing tunable approximations. In particular, we analyzed and enriched the original implementation with tunable software knobs to explore and control the performance-accuracy tradeoffs. 
We modeled time-to-solution of the virtual screening task as a function of software knobs, input data features, and available computational resources. Therefore, the application can autotune its configuration according to a user-defined time budget.
We used a Mini-App derived by LiGenDock -- a state-of-the-art molecular docking application -- to validate the proposed approach.  
We run the enhanced Mini-App on an HPC system by using a very large database of pockets and ligands.
The proposed approach exposes a time-to-solution interval spanning more than one order of magnitude with accuracy degradation up to 30\%, more in general providing different accuracy levels according to the needs of the virtual screening campaign.

\end{abstract}

\begin{IEEEkeywords}
Approximate Computing, Application Autotuning, Molecular Docking, Virtual Screening.
\end{IEEEkeywords}
}

\maketitle

\newcommand{\code}[1]{\texttt{#1}}
\newcommand{\param}[1]{\textsc{#1}}

\IEEEpeerreviewmaketitle

\section{Introduction}
\label{sec:intro}

The end of Dennard scaling pushed towards new sources of computing efficiency. In this direction, one of the most recent and promising trends is the introduction of approximations. Approximate computing wants to save unnecessary computation efforts while finding results whose quality is considered good enough by the end-user.
The reasons behind the increasing popularity of this trend lie in the changing nature of the workloads driving the computing demand~\cite{AxDAC15,Chippa:2013:ACI:2463209.2488873}.
Data mining, computer vision, machine learning, audio and video processing, dynamic simulations, and other classes of applications exhibit an intrinsic resilience to approximations.
In these cases, the purpose of the computing system is not to calculate a precise numerical solution. Indeed, the \emph{correctness} of such systems is defined as a constraint on a metric that represents the quality of the results in the continuum, i.e. a \emph{good enough} result.
The behavior of these applications is mainly related to the type of input/output data, the absence of a unique result, the complexity of finding the correct answer, and the way used to find the results.
More specifically, approximate computing covers the following cases:
\begin{enumerate*}[label=(\roman*)]
\item Applications designed to deal with noisy input data (such as data coming from sensors), or applications whose end-users are not capable to differentiate among precise results (such as human eyes or ears);
\item A unique solution does not exist and a range of solutions are equally acceptable, e.g. recommendation systems or web searches;
\item Although a unique solution exists, the search algorithm is not guaranteed to find it due to the complexity of the solution space (such as in the cases of heuristic searches and machine learning applications);
\item Applications based on patterns, such as iterative-refinement, which guarantee that the precise version is eventually reached (i.e. Montecarlo methods).
\end{enumerate*}

Approximate computing broadly refers to techniques that exploit the intrinsic resilience of applications to realize improvements in efficiency at different layers of the computing stack \cite{Mittal16AxSurvey}. However, in this paper, we focus on the context of software approximations.

The quest for computational effectiveness is not only pushing new software domains towards HPC infrastructures, but also towards legitimizing approximate results in domains traditionally bound to absolute exactness.
One example comes from the pharmaceutical domain and, more specifically, from the drug discovery process. 
The problem consists of finding new drugs starting from a huge exploration space of possible molecules. 
Drug discovery is typically composed of several steps, one of those is the virtual screening procedure to find the most promising molecules to interact with a target protein.
The intrinsic characteristics of the virtual screening let us to classify it as a heuristic search, which is one of the aforementioned categories well-suited for approximate computing.

In this context, the domain-expert belonging to a pharmaceutical company defines the input set for the virtual screening application and the time-to-solution allocated to the job.
Typically, the task of defining the input set fits well the domain-knowledge of the company, while the task of defining the time-to-solution is more complex.
Usually, the company defines a time budget set by the experiment cost and it relies on a domain-expert to tune the size of the molecule database accordingly.
This workflow limits the exploration space without any guarantee neither to find a global optimum nor to find a good local optimum.
Therefore, a reduction in the time spent on evaluating a single pair molecule-protein enables the end-user to explore a larger set of candidates, thus increasing the probability to find an interesting solution.

In this paper, we focus on \MiniApp{}, a molecular docking Mini-App for High-Performance Computing (HPC) systems, based on the LiGenDock module~\cite{beato2013use}, to capture the geometrical features only.
More in general, Mini-Apps represent an important aid for evaluating the architecture and algorithm design space to be explored in the early stages of the code development.
\MiniApp{} attempts to capture key computation kernels of the molecular docking application for the drug discovery implemented in LiGenDock and exploiting only geometrical features.
By developing \MiniApp{} in parallel with the new version of LiGenDock, application developers can work with system architects and domain-experts to evaluate alternative algorithms to better satisfy the end-user constraints or to better exploit the architectural features.
\MiniApp{} enables us a faster performance analysis and the optimization of the key kernels.

The main goal of the proposed approach is to provide tunable approximations to explore performance-accuracy tradeoffs during the docking task, and autotuning \cite{AutotuningInHPC} to support the optimization phase.
In this paper, tradeoffs are enabled enhancing \MiniApp{} by exposing five tunable software knobs.
Then, we autotune them with the support of a predictive model to control time-to-solution in the virtual screening task.
%
In particular, the main contributions of this paper can be summarized as follows:
\begin{itemize}
    \item We analyze \MiniApp{} to properly introduce approximate computing techniques on the most significant kernels;
    \item We enable performance/accuracy tradeoffs by exposing tunable software knobs to drive the \MiniApp{} approximations;
    \item We present the \MiniApp{} performance model based on the exposed software knobs and input data features for estimating time-to-solution in a virtual-screening process;
    \item We enhanced \MiniApp{} with an autotuning layer to satisfy the user-defined time budget according to the workload characteristics.
\end{itemize}

The paper is organized as follows.
In Section \ref{sec:relate}, we present the background on drug discovery, virtual screening and molecular docking needed to understand the context of the work.
Section \ref{sec:method} introduces the proposed methodology for parameterizing, approximating and autotuning the application implementation.
We introduce the experimental setup in Section \ref{sec:experiment_setup},
while in section \ref{sec:experimental_result} we quantitatively evaluate the proposed approach.
Section \ref{sec:relate} presents the related work on molecular docking techniques, approximation methods and autotuning.
Finally, Section \ref{sec:conclusion} summarizes the overall findings and concludes the paper.

\section{Background}
\label{sec:Background}

The goal of a drug discovery process is to find new drugs starting from a huge exploration space of candidate molecules.
Typically, this process involves several \textit{in vivo}, \textit{in vitro} and \textit{in silico} tasks ranging from chemical design to toxicity analysis.
Molecular docking represents one stage of this process \cite{Lionta2014, Beccari2017NovelSP}.
It aims at estimating the three-dimensional pose of a given molecule, named \textit{ligand}, when it interacts with the target protein.
The ligand is much smaller than the target protein, therefore we focus on a small region of the target protein (or receptor), named \textit{pocket} (or binding site).
Given the three-dimensional pose of the ligand within the pocket, we can estimate the strength of the chemical and physical interactions between the ligand and the pocket by computing a geometric fitting score.

The evaluation of the pose of each ligand is independent of the evaluation of all the other candidates.
In drug discovery, being the number of ligands to be analyzed above the billion of units, the problem can be considered embarrassingly parallel.
However, to find the three-dimensional pose of the ligand when it interacts with the pocket, we have to deal with a large number of degrees of freedom.
While the target pocket is represented as a rigid structure, the ligand is represented as a flexible set of atoms bound together by chemical bonds, i.e. sharing electron pairs between atoms (covalent bond).
From a purely geometrical point of view, it is possible to identify a subset of bonds -- named \textit{rotamers} -- which can split the ligand into two disjoint non-empty fragments when they are removed.
We can independently rotate each of those fragments without altering the chemical connectivity of the ligand.
Therefore, we have to consider changes in the shape of the ligand that can be obtained through the rotation of all its rotatable fragments.

Evaluating the chemical and physical interactions between the ligand and the pocket is a computationally-intensive problem, therefore state-of-the-art approaches~\cite{Kitchen2004,LYNE20021047,doi:10.1021/jm010494q} suggest to separate the pose prediction task from the virtual screening task.
The pose prediction task focuses on providing the best pose for a given ligand in a given binding site, whereas the virtual screening task aims at selecting in a huge database of candidates a small set of promising ligands which best fit the given binding site.
The structure of the two tasks are very similar to each other.
Several industrial applications~\cite{beato2013use,friesner2004glide} provide both functionalities in the same software module.
A remarkable difference between the pose prediction and the virtual screening task lies in the approach used for the estimation of the chemical and physical interactions between the ligand and the pocket.
It is possible to estimate such interactions by using either a geometrical or a pharmacophoric approach.
The geometrical approach estimates the ligand-pocket interactions by using information related only to the shape and volume, while the pharmacophoric approach evaluates the actual chemical and physical interactions.
The latter approach is the most computational-intensive one and it is regularly exploited on the pose prediction task.
Although the best solution according to a pharmacophoric score implies a good geometrical score, there are no guarantees on the opposite case.
The best solution according to a geometrical score might be either a non-valid solution or a poor solution from a pharmacophoric perspective.
Therefore, there is always the need to apply a pharmacophoric docking process.
The geometrical approach is used for filtering all the solutions that cannot geometrically fit the target pocket.

In this paper, we exploit the application parameterization targeting approximation and autotuning for controlling the time-to-solution in a virtual screening campaign. This is done according to the target database of molecules (i.e. ligands) to be docked and to the characteristics of the protein binding site (i.e. pocket).


\section{Methodology}
\label{sec:method}

This section first introduces \MiniApp{} by describing the algorithm and highlighting the application hot spots.
Then, we describe the functional analysis used to drive the approximations of the application enabling the accuracy-throughput tradeoffs.
Finally, we describe how end-users might leverage the exposed tradeoff for autotuning the application according to the time-to-solution constraints and to the workload characteristics.

\subsection{Application Description}
\label{subsec:app_descr}
\newcommand{\kernel}{\code{MatchProbesShape}\xspace} 

In the context of the LiGen toolflow~\cite{beccari2013ligen}, LiGenDock~\cite{beato2013use} is the module for docking one or more ligands into a target protein.
It can be used for both the pose prediction and the virtual screening tasks.
It exploits chemical and geometrical features to dock the ligand through an iterative algorithm.
In particular, LiGenDock uses chemical features to set the initial pose of the ligand and to drive the docking process at each iteration.
However, it also uses geometrical features to optimize the pose of the ligand at each iteration, by taking into account all the degrees of freedom of the problem space.

The optimization of the ligand pose is the most computationally-intensive part of LiGenDock during the virtual screening task. 
\MiniApp{} includes all the functionalities of LiGenDock that optimize the ligand pose by exploiting the geometric approach.
\MiniApp{} takes as input the target pocket and a database of ligands, and it generates as output the score of each pocket-ligand pair, after optimizing the pose of the ligand.

\MiniApp{} performs the virtual screening task by using the geometric approach.
It estimates the pocket-ligand interactions with the similarity between the shape of the ligand and the three-dimensional shape of the pocket in PASS format \cite{brady2000fast}, which is generated by LiGen PASS~\cite{beccari2013ligen}.
Actually, \MiniApp{} scores each ligand by using the \emph{overlap score} function.
The overlap score, as defined in Equation \ref{eq:overlap-score}, is the reciprocal of the minimum square distance between the ligand and the pocket:
\begin{equation} \label{eq:overlap-score}
o = \frac{l}{\sum\limits_{i=0}^l \min\limits_{j=0}^p d^2(L[i],P[j])}
\end{equation}
where $o$ is the overlap score, $l$ is the number of atoms in the ligand $L$, $p$ is the number of 3D points in the pocket $P$, and $d^2$ represents the squared distance between the $i$-th atom of the ligand and the $j$-th point of the pocket.
Higher overlap means better geometric compatibility between the pocket and the ligand. 

\begin{figure}[t]
	\centering
    \includegraphics[height=1.0\columnwidth, angle=-90]{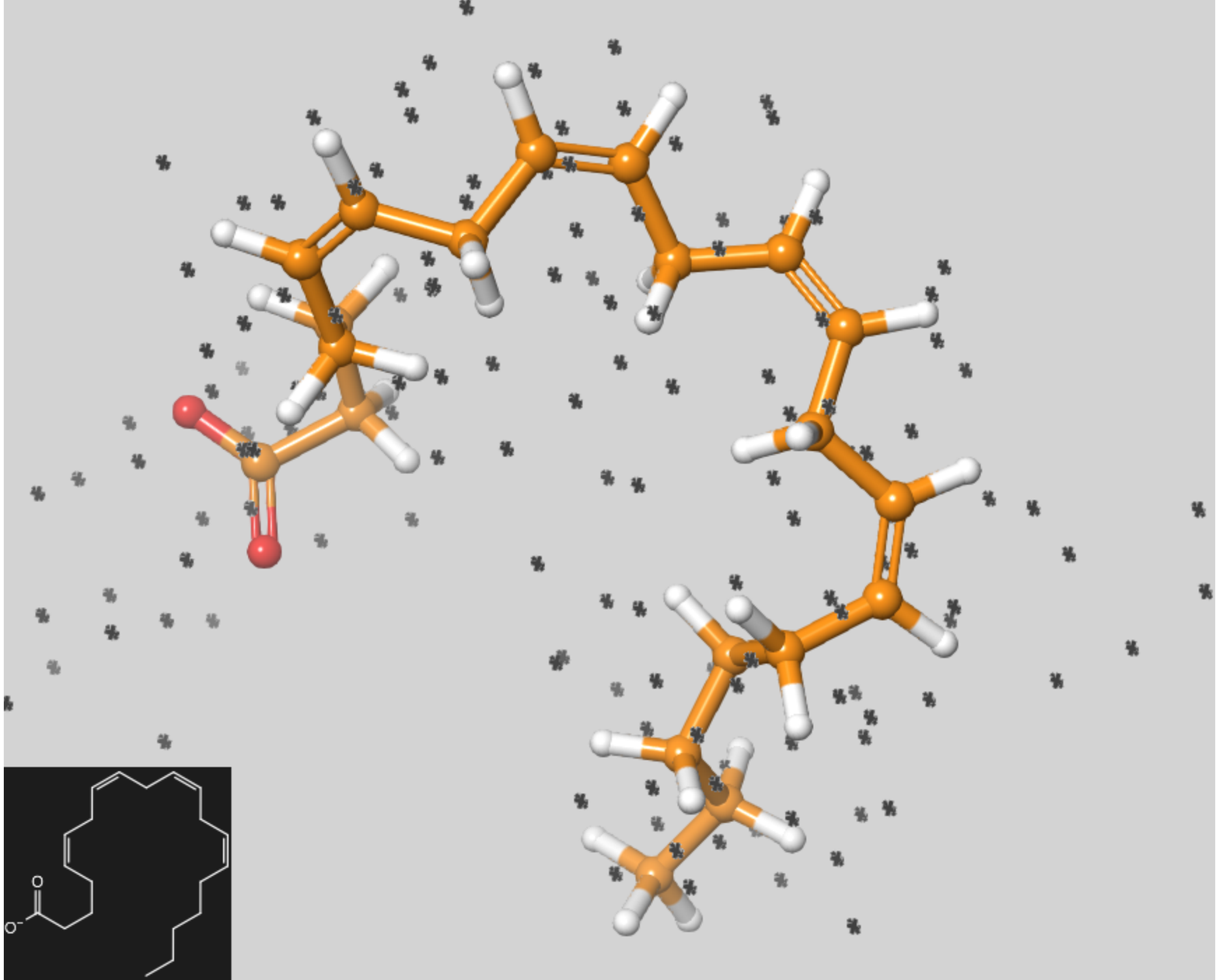}
\caption{3D visualization of a docked ligand (connected structure) inside a PASS version of the target pocket (dark spots). }
\label{fig:pocket-ligand}
\end{figure}

\prettyref{fig:pocket-ligand} shows an example of docking a ligand inside a pocket (i.e. 1cvu \cite{Berman00theprotein}). The ligand structure is clearly visible in the 3D-image and its planar representation is highlighted in the bottom left corner. The larger bubbles are the atoms $L$ of the ligand, while the connections between atoms are the molecule bonds. The dark spots in \prettyref{fig:pocket-ligand} are the points $P$ representing the PASS version of the target pocket.
These points are the centre of the spheres used to model the binding site.

\subsection{Analysis of \MiniApp{} }
\label{subsec:kernel_analysis}

\MiniApp{} is designed to target an HPC platform and it leverages machine-level parallelism by using the MPI master/slave paradigm.
In particular, the master process reads the ligands' input database and dispatches a ligand to be evaluated to any available slave.
Each slave docks geometrically the ligand in the binding site of the target molecule, which is constant for all the evaluated ligands.
Finally, the slave computes the final geometrical score, sends its value to the master process, and waits for the next ligand to be evaluated. 
\MiniApp{} (similar to the original LiGenDock) leverages the parallelism at data-level, where each slave evaluates a pocket-ligand pair independently.
Given the huge number of ligands to be evaluated, the benefit of this approach is twofold.
On one side, the docking algorithm is serial, therefore it avoids any synchronization mechanism in the evaluation of the ligand-pocket pair.
On the other side, the application can efficiently leverage the embarrassingly parallel nature of the problem at a high-level.

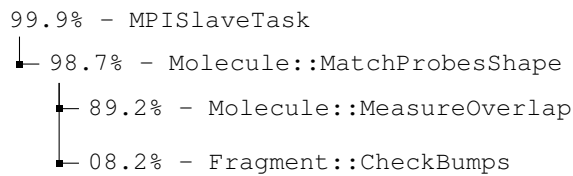
\begin{figure}[t]
\centering
\begin{forest}
  for tree={
    font=\ttfamily,
    grow'=0,
    child anchor=west,
    parent anchor=south,
    anchor=west,
    calign=first,
    edge path={
      \noexpand\path [draw, \forestoption{edge}]
      (!u.south west) +(7.5pt,0) |- node[fill,inner sep=1.25pt] {} (.child anchor)\forestoption{edge label};
    },
    before typesetting nodes={
      if n=1
        {insert before={[,phantom]}}
        {}
    },
    fit=band,
    before computing xy={l=15pt},
  }
[99.9\% - MPISlaveTask
	[98.7\% - Molecule::MatchProbesShape
		[89.2\% - Molecule::MeasureOverlap]
        [08.2\% - Fragment::CheckBumps]
	]
]
\end{forest}
	\caption{Application Call Graph profile. Functions taking less than 2\% of the overall execution time are omitted}
    \label{fig:gprof_callgraph}
\end{figure}

We profiled the application to identify the critical sections of the code by using \textsc{gprof}\footnote{GNU gprof \url{https://sourceware.org/binutils/docs/gprof/}}.
\prettyref{fig:gprof_callgraph} shows the Call Graph report: It groups the individual functions by the caller.

A large fraction of the execution time of the application is spent on \kernel.
This kernel is responsible for the optimization of the shape of the ligand by using a steepest descent algorithm to deal with all the internal degrees of freedom of the ligand.
In this paper, we focus on the introduction of possible software knobs through approximation techniques to tune the time-to-solution of this functionality.

\begin{algorithm}[t]
 \label{alg:MiniApp_kernel}
 \LinesNumbered
 \KwData{the pocket and the 3D structure of the ligand}
 \KwResult{the overlap score of the ligand }
 get the list of rotamers\; \label{line:get_rot_bonds}
 \ForEach{rotamer}{ \label{line:outer_loop_begin}
  grow the right and left fragment\; \label{line:grow_segm}
  \For{angle in 0-360 degrees with step 1 degree}{ \label{line:rotation1_begin}
   rotate left fragment to $angle$\;  \label{line:inner_rotate_1}
   \If{the ligand shape is feasible }{ \label{line:check_1}
   	 measure the overlap of the ligand\;  \label{line:measure_overlap_1}
   	 check if the overlap is improved
   } \label{line:check_1_end}
  }\label{line:rotation1_end}
  set the left fragment to best angle found\; \label{line:set_best_left}
  \For{angle in 0-360 degrees with step 1 degree}{ \label{line:rotation2_begin}
   rotate right fragment to $angle$\; \label{line:inner_rotate_2}
   \If{the ligand shape is feasible }{ \label{line:check_2}
     measure the overlap of the ligand\; \label{line:measure_overlap_2}
   	 check if the overlap is improved
   } \label{line:check_2_end}
  } \label{line:rotation2_end}
  set the right fragment to best angle found\; \label{line:set_best_right}
 } \label{line:outer_loop_end}
 \Return{the overlap score of the ligand\;}
 \vspace{1em}
 \caption{Pseudo-code of the \kernel kernel, which changes the shape of the ligand to maximize the overlap score.}
\end{algorithm}

\prettyref{alg:MiniApp_kernel} shows the pseudocode of the target kernel.
First, the algorithm identifies the set of \emph{rotamers} (line~\ref{line:get_rot_bonds}), thus selecting all the possible sources of flexibility in the ligand shape.
Then, it iterates over the set of these bonds to find the best shape of the ligand (lines from~\ref{line:outer_loop_begin}~to~\ref{line:outer_loop_end}).
In particular, the body of the algorithm grows a left and a right ligand fragment, with respect the two extremes of the bond (line~\ref{line:grow_segm}).
The left and the right fragments are rotated independently.
The first to be processed is the left fragment.
It is rotated step-by-step up to a full angle (lines from~\ref{line:rotation1_begin}~to~\ref{line:inner_rotate_1}); 
At each step, we check whether the ligand shape is valid.
There is a non-null possibility of internal bumping of the molecule (line~\ref{line:check_1}), which invalidates the shape of the ligand.
If the ligand shape is valid, the overlap score of the ligand is considered during the check for possible improvements (lines from~\ref{line:measure_overlap_1}~to~\ref{line:check_1_end}).
At the end of the 360-degree exploration, we rotate once more the left fragment to match the angle that maximizes the overlap score (line~\ref{line:set_best_left}).
The same procedure is applied to the right fragment (lines from~\ref{line:rotation2_begin}~to~\ref{line:set_best_right}).

The kernel applies the same computation of the left and right fragments of each rotamer. 
For this reason, in the rest of the paper, we do not differentiate the two fragments.

\prettyref{fig:gprof_callgraph} also shows that the computation of the overlap score of each pose (\code{Molecule::MeasureOverlap}) represents the most expensive operation.
The implementation of \code{Molecule::MeasureOverlap} is relatively simple and a lot of effort has already been spent in the past in terms of performance tuning. Our contribution aims at reducing the number of invocations performed by its caller.
In particular, we want to avoid the computations which are very likely to do not lead to any improvement.

\begin{figure}
	\centering
    \includegraphics[width=\columnwidth]{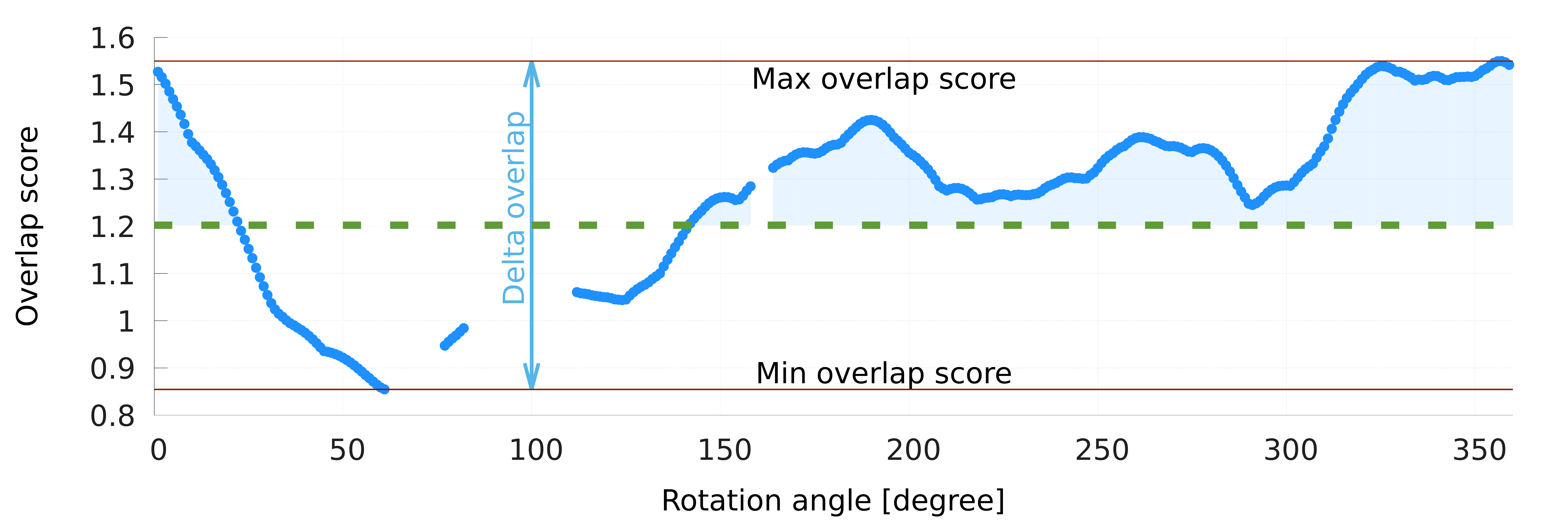}
	\caption{Changes in the overlap score by rotating a fragment of the ligand. The x-axis represents the angle of the rotation, while the y-axis represents the overlap score of the ligand. }
	\label{fig:peak_definition}
\end{figure}

\prettyref{fig:peak_definition} shows an example of how the rotation of a fragment affects the overlap score of the ligand.
The x-axis represents the rotation space, while the blue line shows the overlap score of the ligand according to the position of the fragment. 
The empty spaces are due to the fact that some of the generated poses of the ligands are not valid because of the internal bumps of the ligand atoms.
We define as \emph{delta overlap} the difference between the minimum and maximum overlap of a single rotation.
We define as \emph{peak} the set of contiguous and valid rotation angles whose overlap is higher than 50\% of the delta.

\begin{figure}
	\centering
	\subfloat[Distribution of the delta overlap.]{%
		\label{fig:delta_overlap}%
        \includegraphics[width=1.0\columnwidth]{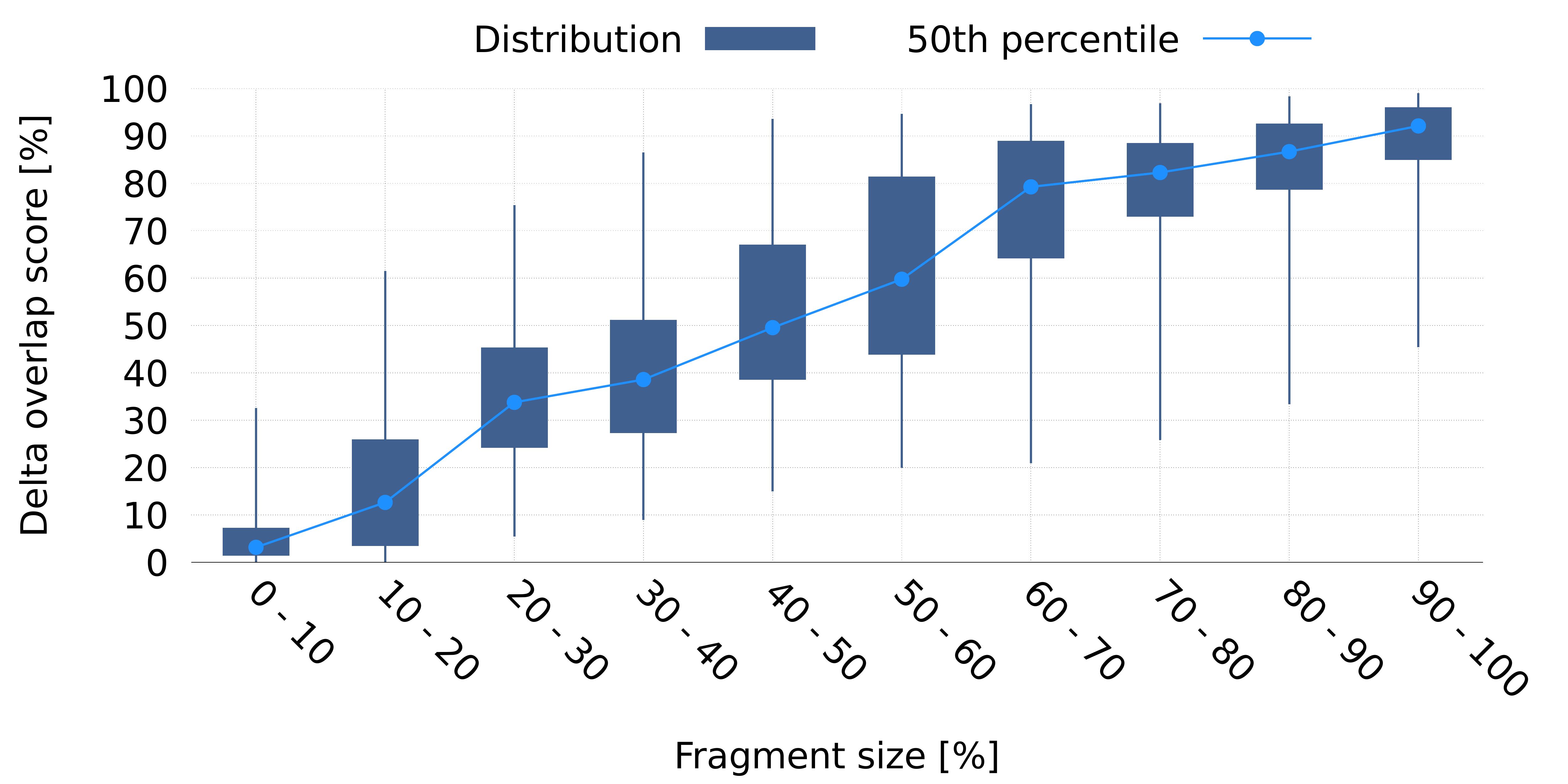}}\\
	\centering
	\subfloat[Distribution of the peak geometry.]{%
		\label{fig:geometry}%
        \includegraphics[width=1.0\columnwidth]{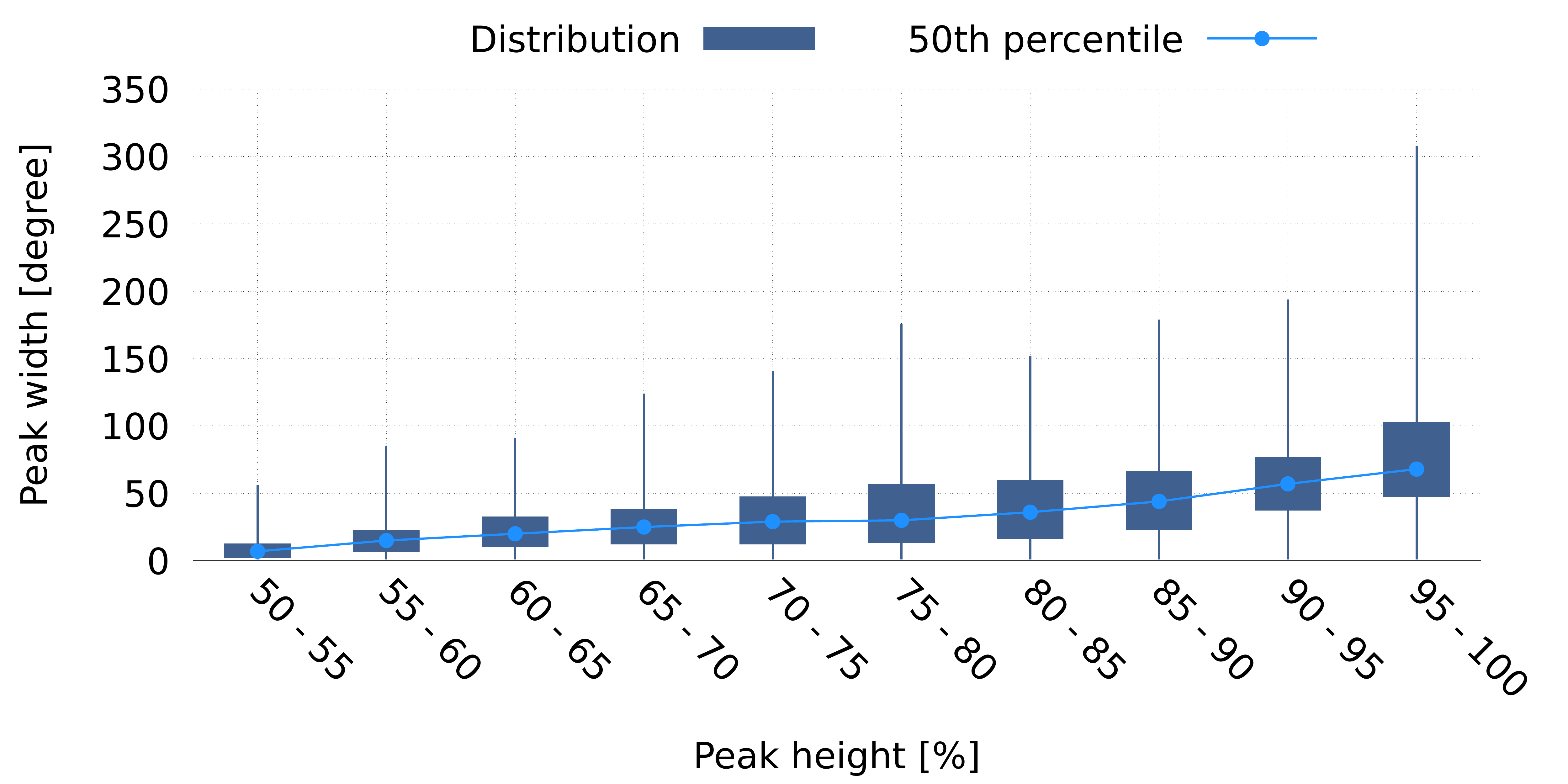}}\\
	\centering
	\subfloat[Distribution of the number of peaks.]{%
		\label{fig:number_rotamer}%
        \includegraphics[width=1.0\columnwidth]{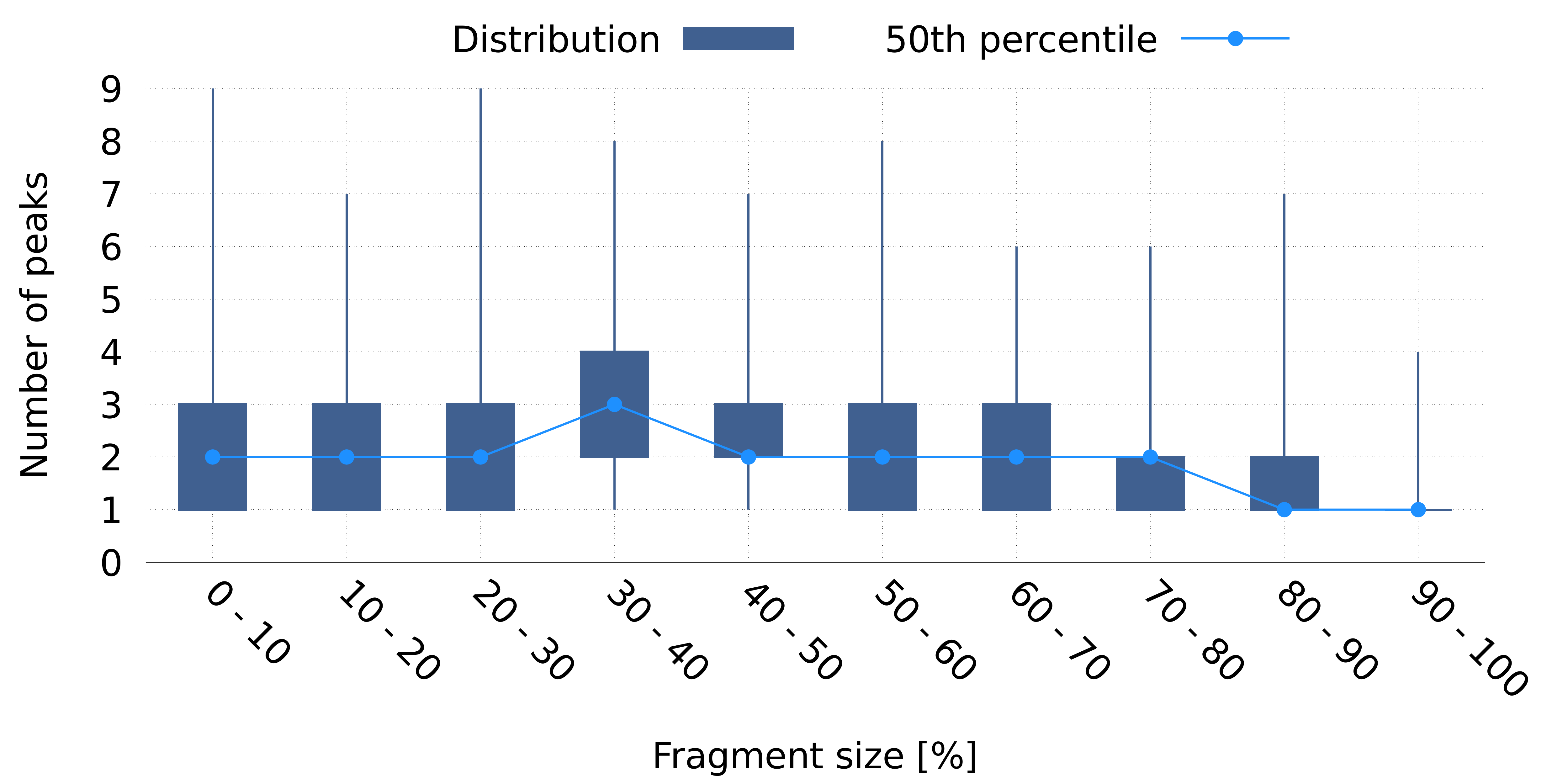}}\\
\caption{Analysis on the peaks of overlap across different fragments.
Each plot shows the minimum value, the 25th, 50th, 75th percentile and the maximum value.}
\label{fig:peak_analysis}
\end{figure}

To further analyze the functional behaviour, we performed an experimental campaign by using a chemical library composed of 113K ligands.
This analysis aims at finding patterns that can be leveraged for reducing the number of evaluations for each fragment rotation, thus creating possible application knobs (see \prettyref{fig:peak_analysis}). 

\prettyref{fig:delta_overlap} correlates the size of a fragment with its impact on the final score of the ligand.
In particular, the x-axis represents the relative size of the fragment with respect to the size of the ligand, while the y-axis represents the delta overlap normalized with respect to the final score of the ligand. 
It is worth noticing that small fragments have small deltas, which means that such fragments usually have a limited impact on the numeric value of the final score of the ligand.

\prettyref{fig:geometry} correlates the width of a \emph{peak} (in degree) with its height normalized with respect to the delta overlap. 
From this plot, we can notice that the peaks that contain the maximum overlap are usually greater than $50$ degrees, while narrow peaks rarely reach the maximum height.
We can conclude that the behaviour of the overlap score is rather smooth because small peaks are also narrow.

\prettyref{fig:number_rotamer} shows on the y-axis the number of peaks which are contained in a fragment by changing the fragment size on the x-axis. We can notice how larger fragments usually have only one peak, while smaller ones tend to have more peaks.

\begin{figure}
	\centering
	\subfloat[\kernel execution time composition.]{%
		\label{fig:execution_time}%
        \includegraphics[width=1.0\columnwidth]{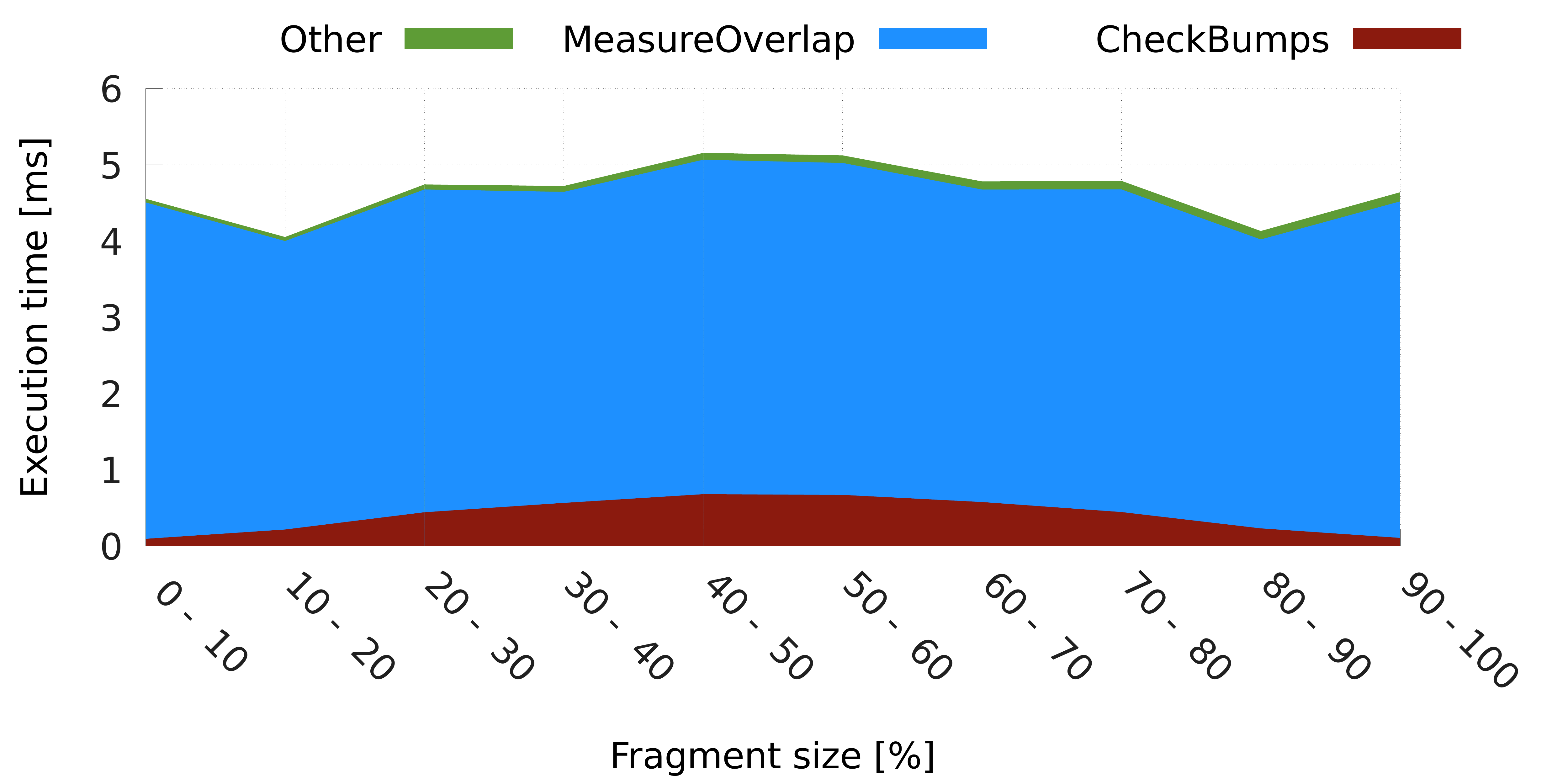}}\\
	\centering
	\subfloat[Frequency distribution of the fragment sizes.]{%
		\label{fig:frequency}%
        \includegraphics[width=1.0\columnwidth]{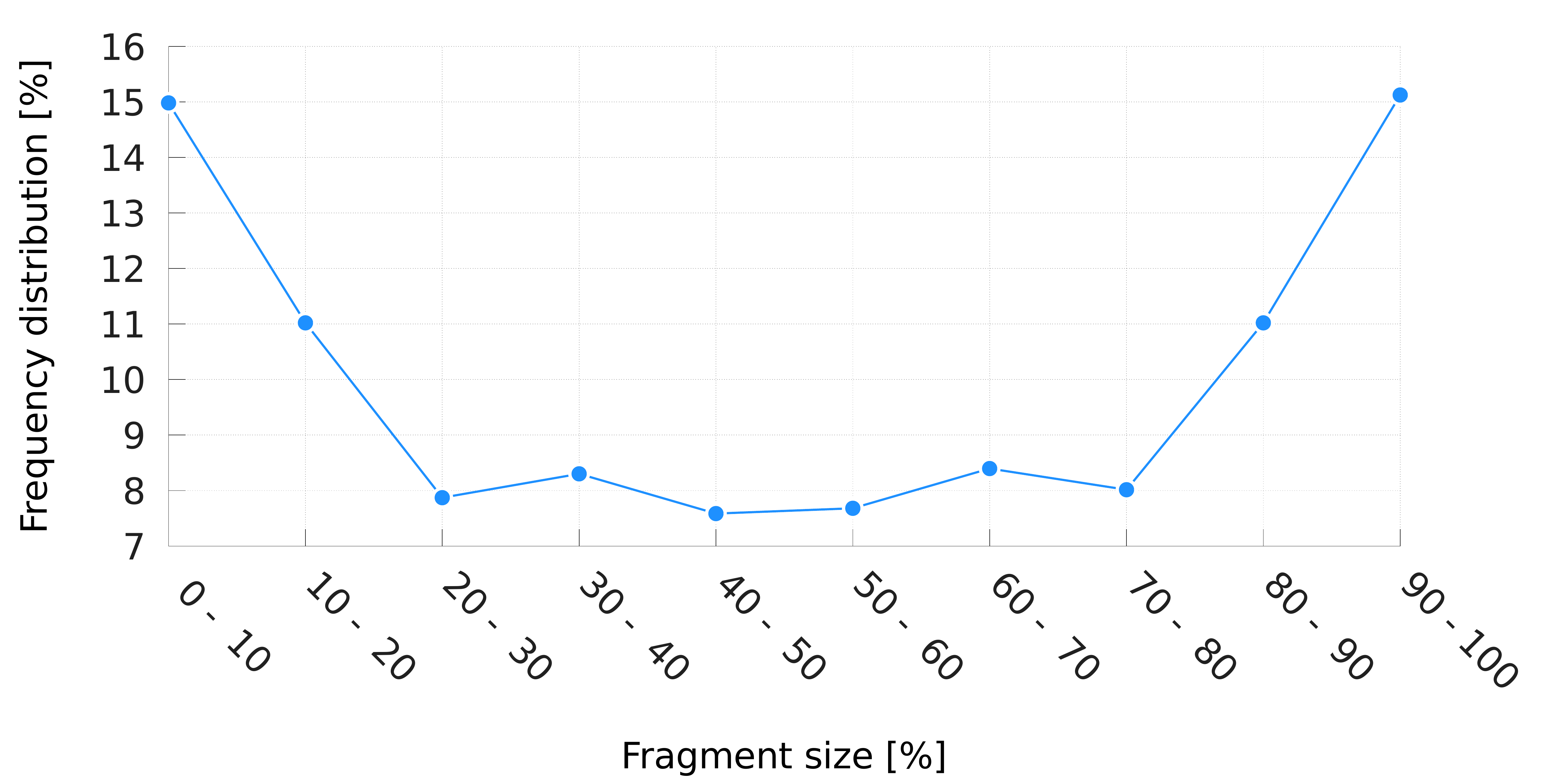}}\\
\caption{Analysis of execution time and frequency distribution of fragments, grouped by their size.}
\label{fig:approx-approac}
\end{figure}

Besides the functional behaviour of the overlap score, \prettyref{fig:execution_time} shows the detailed composition of the time spent by the application to find the best rotation angle of a fragment of the ligand (y-axis) according to the size of the fragment (x-axis).
From the execution time, we highlight the time spent by measuring the overlap score (MeasureOverlap) and the time spent by checking if the evaluated angle is admissible (CheckBumps).
From the plot, we see that computing the overlap score is independent of the size of the fragment.
This result is expected because it involves the evaluation of the whole ligand.

Finally, \prettyref{fig:frequency} shows the frequency distribution of the size of a fragment.
Due to the definition of the ligand database, smaller and larger fragments are slightly more frequent with respect to the other fragment sizes.

\subsection{Exposing Tunable Application Knobs}
\label{subsec:approx}

In the original LiGenDock \cite{beato2013use}, the authors listed several parameters that can alter the behaviour of the docking algorithm.
However, most of them are chemical-specific parameters that do not impact the execution time of \MiniApp{}.
The only exception is a constant parameter which performs loop perforation \cite{Sidir_loop_perf} on the loops that rotates a fragment of the ligand (lines from~\ref{line:rotation1_begin}~to~\ref{line:inner_rotate_1} in \prettyref{alg:MiniApp_kernel}).
In particular, the baseline version uses a step of $1°$, while it is possible to increase the step size to skip iterations, thus reducing the number of evaluations, thus increasing the performance of the application.

In addition to this first step, based on the analysis done in the previous section, we can exploit domain-knowledge to define more aggressive software-knobs to approximate the \MiniApp{} results.
In particular, we know from \prettyref{fig:delta_overlap} that small fragments have a limited impact on the delta overlap.
Therefore, instead of applying a flat loop perforation (as done in the original application), we introduce a \emph{parametric loop perforation} which let us  focus on the most important fragments of the ligand.
Whenever the size of the current fragment is below a given \param{threshold}, we use a coarse-grain rotation step (\param{low-precision step}). Otherwise, we use a fine-grain rotation step (\param{high-precision step}).

Given that \kernel is a greedy algorithm, we might improve the overlap score by repeating the whole procedure, thus considering multiple time each fragment.
In particular, the more we repeat the procedure, the more we increase the probability to find a better pose for the target ligand.
Therefore, we define the tunable software knob \param{repetitions} as the number of times to repeat the procedure.
This step seems to be counter-intuitive, however, we argue that it is better to run more times \kernel with aggressive approximations instead of running it  only once with fewer approximations.

Furthermore, we can extract other important information about the overlap score from the peak analysis discussed in the previous section.
In particular, we can rely on the smoothness of the overlap score through the entire rotation space, which means that each fragment has a limited number of peaks and that the most important peak is usually wide (the median is $68^\circ$).
Therefore, we can further reduce the number of evaluations by introducing an \emph{iterative refinement} technique based on the concept of loop tiling and grid-based approaches.
For each fragment above \param{threshold} we partition the $360^\circ$ rotation space into several tiles of fixed-size $x$.
Then, we peel and evaluate only one element for each tile (the central one).
In the next iteration, we evaluate only the tile corresponding to the most promising peeling element by using \param{high-precision step}.
Given this policy, the number of evaluated rotations ($y$) is a function of the tile size ($x$) and \param{high-precision step}, as described in \prettyref{eq:num_iterations}.
\begin{equation} \label{eq:num_iterations}
y = \frac{360^\circ}{x} + \frac{x}{\param{high-precision step}}
\end{equation}
We are interested in minimizing the number of pose evaluation while preserving a high probability to identify the most important peak.
The minimization of \prettyref{eq:num_iterations} has a unique solution, its value $\hat{x}$ is defined in \prettyref{eq:min_num_iterations}.
\begin{equation} \label{eq:min_num_iterations}
\hat{x} = 6*\sqrt[]{10}*\sqrt[]{\param{high-precision step}}
\end{equation}

\begin{figure}[t]
	\centering
    \includegraphics[width=1.0\columnwidth]{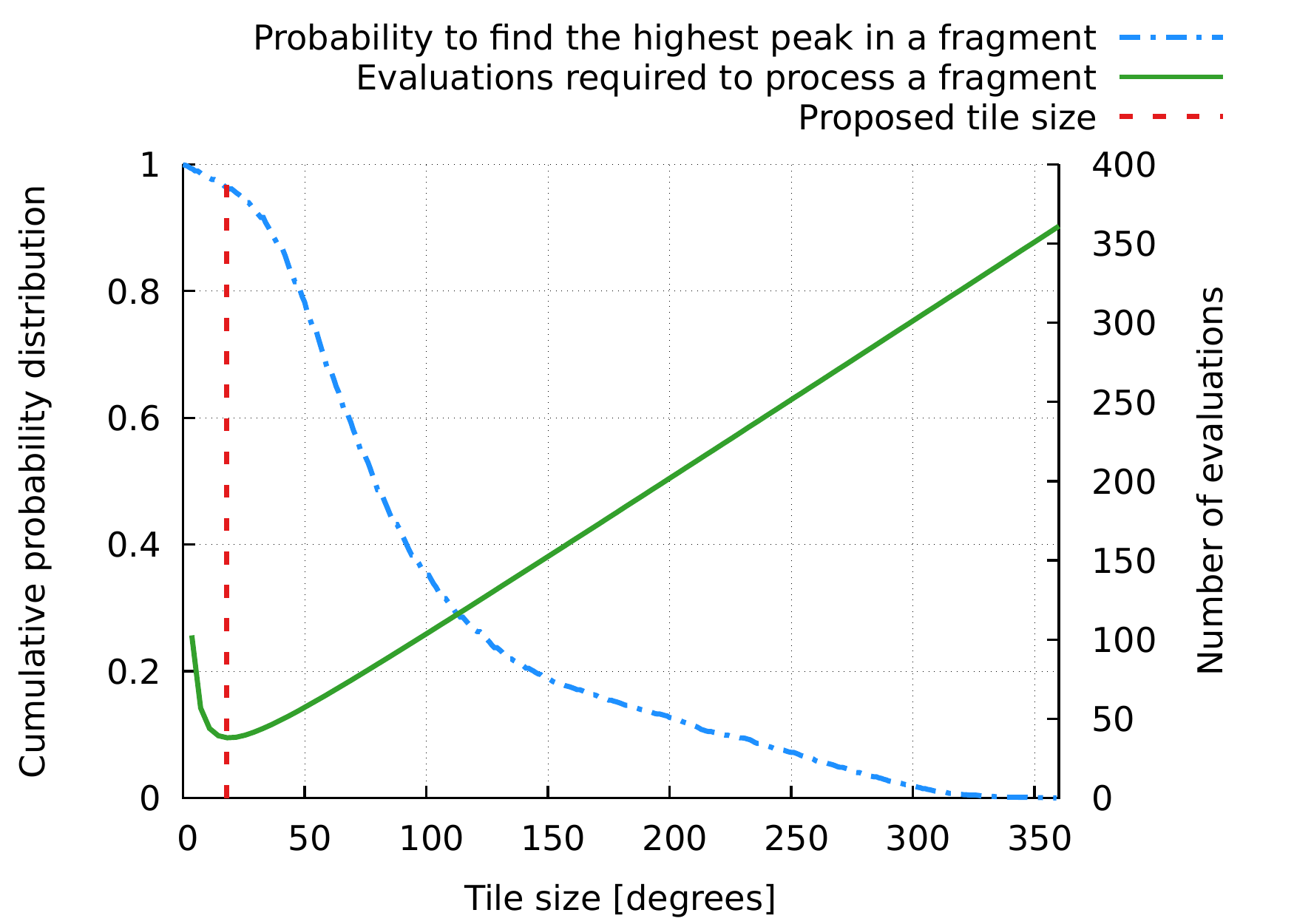}
\caption{For each tile size (x-axis), the relation between the number of evaluated rotations (y2-axis) and the probability that the width of the best peak is greater than the given size (y1-axis). }
\label{fig:tile-size}
\end{figure}

For example, if we set \param{high-precision step} at the same accuracy of the original algorithm ($1^\circ$), the optimal tile size is $18^\circ$, which means that we have a high probability to identify the most important peak with the peeling element. 
In particular, \prettyref{fig:tile-size} shows, for each tile size (x-axis), the probability that the width of the most important peak is greater than the evaluated tile size (y1-axis, blue line) and the number of evaluated iterations (y2-axis, green line).
The red line highlights the optimal value.
As a consequence of \prettyref{eq:min_num_iterations}, we observe that an increment of \param{high-precision step} implies an increment in the value of the optimal tile size and a decrement of the probability of finding the best peak.

To summarize, starting from the original algorithm described in \prettyref{alg:MiniApp_kernel}, we have introduced five tunable software-knobs (\param{high-precision step}, \param{low-precision step}, \param{threshold}, \param{repetitions} and \param{enable refinement}) to approximate the application by reducing the number of ligand evaluations.
The main idea is to spend more time when the computation is more promising. 

\begin{algorithm}[t]
 \label{alg:MiniApp_kernel_final}
 \LinesNumbered
 \KwData{the pocket and the 3D structure of the ligand}
 \KwResult{the overlap score of the ligand }
 get the list of rotamers\;
 \For{the number of \param{repetitions}}
 {
 	\label{line:repetitions}
 	\ForEach{rotamer}
    {
    	grow the right and left fragment\;
        \If{relative size of left fragment $\leq$ \param{threshold}}
        {
        	\label{line:left_start}
            place the left fragment in the best angle found with step \param{low-precision step}\;
            \label{line:coarse}
        }
        \Else
        {
        	\If{\param{enable refinement}}
            {
            	\label{line:fine_start}
            	evaluate the peeling element for each tile\;
                \label{line:peeling}
                place the left fragment in the best angle found in the best tile using step \param{high-precision step}\;
                \label{line:tile}
            }
            \Else
            {
            	place the left fragment in the best angle found with step \param{high-precision step}\;
                \label{line:fine}
            }
            \label{line:fine_stop}
        }
        \label{line:left_stop}
        
        \If{relative size of right fragment $\leq$ \param{threshold}}
        {
        	\label{line:right_start}
            place the right fragment in the best angle found with step \param{low-precision step}\;
        }
        \Else
        {
        	\If{\param{enable refinement}}
            {
            	evaluate the peeling element for each tile\;
                place the right fragment in the best angle found in the best tile using step \param{high-precision step}\;
            }
            \Else
            {
            	place the right fragment in the best angle found with step \param{high-precision step}\;
            }
        }
        \label{line:right_stop}
    }
 }

 \Return{the overlap score of the ligand\;}
 \vspace{1em}
 \caption{The tunable pseudo-code of the \kernel kernel.}
\end{algorithm}

In particular, \prettyref{alg:MiniApp_kernel_final} shows the parametric algorithm of \kernel.
All the original algorithm is contained in the outer loop (\prettyref{line:repetitions}) which repeats the pose optimization according to \param{repetitions}.
The optimization of the pose of left fragment is described between \prettyref{line:left_start} and \prettyref{line:left_stop}.
According to the relative size of the fragment and to \param{threshold} (\prettyref{line:left_start}), we run either a coarse grained exploration by using \param{low-precision step} (\prettyref{line:coarse}) or a fine grained exploration (lines \ref{line:fine_start}-\ref{line:fine_stop}).
In the latter case, we either perform a two-step optimization by using iterative refinements or we perform a flat exploration using \param{high-precision step}, according to \param{enable refinement}. 
The two-step optimization evaluates the peeling elements of the rotation (\prettyref{line:peeling}) and then it refines the exploration of the most promising tile by using \param{high-precision step} (\prettyref{line:tile}).
Due to the symmetry of the problem, the same procedure is applied to the right fragment (lines \ref{line:right_start}-\ref{line:right_stop}).

\subsection{Application Autotuning}
\label{subsec:application_autotuning}

The software knobs defined in the previous section aim at reducing the exploration space of ligand poses, decreasing the time-to-solution of the application and the accuracy of the results as well. However from the end-user point of view, a manual selection of the application configuration it is a non-trivial task.
Application autotuning \cite{AutotuningInHPC} is a well-known field in literature and there are several available tools \cite{ORIO, ACTIVEARMONY, PTF} that can select the most suitable configuration according to application requirements.
In this section, we explain how tuning can be automatically done by the application itself according to an execution time budget allocated by the end-user.
We used the mARGOt \cite{gadioli2018margot} framework to select the configuration of the software knobs that maximizes the accuracy of the result given the time budget.

To select the most suitable configuration, the autotuner should be able to predict the performance of the configurations for the actual input \cite{PETABRICKS_input}.
As the accuracy is platform-independent and it is used to sort the configurations in terms of software knobs, it is possible to run an error profiling campaign only once, averaging the results over a representative set of pockets and ligands.

To complete the execution of the application in the given time budget, we have to estimate the time-to-solution once the target architecture and the actual input dataset (pocket and ligands database) are given.
Being the target problem embarrassing parallel, without the need of synchronization, the overhead of the MPI environment is negligible even scaling over a large set of nodes.
Therefore, assuming homogeneous resources, we predicted the time-to-solution of the serial case, then we re-calculate according to the allocated resources. 

Considering only a configuration for the software knobs, it is possible to use input data features to estimate the time-to-solution of the given input.
To this end, we modelled the entire database as a set of ligands with the same \emph{average} data features. 
In particular, we used a multivariate linear regression with interaction to estimate the time-to-solution $t_{la}$ for the average ligand. The vector of predictors $\overline{x}$ is composed of the number of 3D points of the pocket $xp_p$, the average number of atoms in a ligand $xl_a$, the average number of rotamers in a ligand $xl_r$, and all the possible interactions among them (i.e. $xp_p \cdot xl_a$, $xp_p \cdot xl_r$, $xl_a \cdot xl_r$, and $xp_p \cdot xl_a \cdot xl_r$). 
Thus, the target model is simply composed of $t_{la} = \overline{\alpha} \cdot \overline{x} + \beta$, where $\overline{\alpha}$ is the vector of predictor coefficient, while $\beta$ is the intercept.

To generalize the approach, we considered the parameters of the regression as a function of the proposed software knobs, given that the impact of the data features on the execution time is strongly dependent on the configuration.
By using this information, we can build a model to estimate the time-to-solution according to \prettyref{eq:linear_regression}, where $\overline{k}$ is the vector of software knobs and $\nu$ is the number of ligands to dock, in the input database.

\begin{equation} \label{eq:linear_regression}
t = \nu \cdot (\overline{\alpha}(\overline{k}) \cdot \overline{x} + \beta(\overline{k}))
\end{equation}

Differently from the accuracy characterization, the performance model should be trained every time the computing platform is changed. However, in both cases, the experiment described in \prettyref{subsec:datadependence} suggests that a small database of ligands is enough to define the accuracy-performance behaviour.

To recap, we enhanced the original algorithm of the application by exposing software knobs that enable performance-accuracy trade-offs.
We used an application autotuner to configure automatically the application according to simple user-oriented parameters: the number of available computational resources and the available time-budget. The data features of the actual input can be either included by the user or directly extracted by a preliminary input data analysis.

\section{Experimental Setup}
\label{sec:experiment_setup}

To assess the benefits of the approximation techniques described in this paper, the experimental setup consists of the data sets used in the experiments, the metrics of interest and the target platform to execute the application.

\subsection{Data Sets}

To evaluate the functional and extra-functional performance of the proposed approximation techniques, we used a database of ligands composed of 113K ligands.
The molecules are different in terms of the number of atoms (between $28$ and $153$) and rotamers (between $2$ and $53$).
We used 6 pockets protein-ligand complexes derived from the RCSB Protein Data Bank (PDB) \cite{Berman00theprotein}: 1b9v, 1c1v, 1cvu, 1c2, 1dh3, 1fm9. In particular, the PASS \cite{brady2000fast} version of the pockets has also been used together with the database of ligands.
Differently from the classical grid representation of the pocket, the PASS (Putative Active Sites with Spheres) version uses spheres to represent binding sites. This solution has been widely used in the context of fast docking \cite{brady2000fast}. 

\subsection{Metrics of Interest}
To measure the performance of \MiniApp{}, we considered its throughput and the time-to-solution.
In particular, the application throughput is defined as the number of ligand's atoms processed in one second, while the time-to-solution is the time taken by the application to elaborate the input.

We used the metric \textit{overlap degradation} to quantify the mean loss of accuracy introduced by the approximation techniques with respect to the baseline, which is the configuration that leads, on average, to the better overlap score: $\param{high-precision step}=1^\circ$, $\param{threshold}=0$, $\param{repetitions}=3$ and $\param{enable refinement}=false$.
The \textit{overlap degradation} is defined as described in \prettyref{eq:error_degradation},
\begin{equation} \label{eq:error_degradation}
score_{degradation} = ( 1 - \frac{overlap_{approx}}{overlap_{original}}) \times 100
\end{equation}
where $overlap_{approx}$ is the mean overlap score of the top $1\%$ ligands of the evaluated configuration, while $overlap_{original}$ is the mean overlap score of the top $1\%$ ligands of the non-approximated version of the application (baseline).

\subsection{Target Platform}
\label{subsec:target}
The platform used to execute the experiments is based on dedicated supercomputer NUMA nodes featuring two Intel Xeon E5-2630 V3 CPUs (@2.8 GHz) with 128 GB of DDR4 memory (@1866 MHz) on a dual channel memory configuration.
The experiments are executed by using the GALILEO platform located at CINECA supercomputing center\footnote{https://www.cineca.it/en}. 

\section{Experimental Results}
\label{sec:experimental_result}
In this section, we evaluated the benefits of the proposed methodology by using four different experiments.
Being a data-dependent application, the first experiment aims at assessing data sensitivity by changing the number of ligands used for evaluating a configuration.
The second experiment assesses the benefits of applying the approximation techniques with respect to the original version of the application. We show the obtained accuracy-performance tradeoffs for a virtual screening procedure and we evaluate the effects of the overlap degradation on a single ligand docking procedure. The third experiment validates the accuracy of the time-to-solution model.
Finally, the fourth experiment wants to prove the benefits of the proposed approach for the end-user on two different application scenarios.

\subsection{Data Dependency Evaluation}
\label{subsec:datadependence}

\begin{figure}
	\centering
	\subfloat[Throughput per process]{%
		\label{fig:galileo_throughput_plot}%
        \includegraphics[width=1.0\columnwidth]{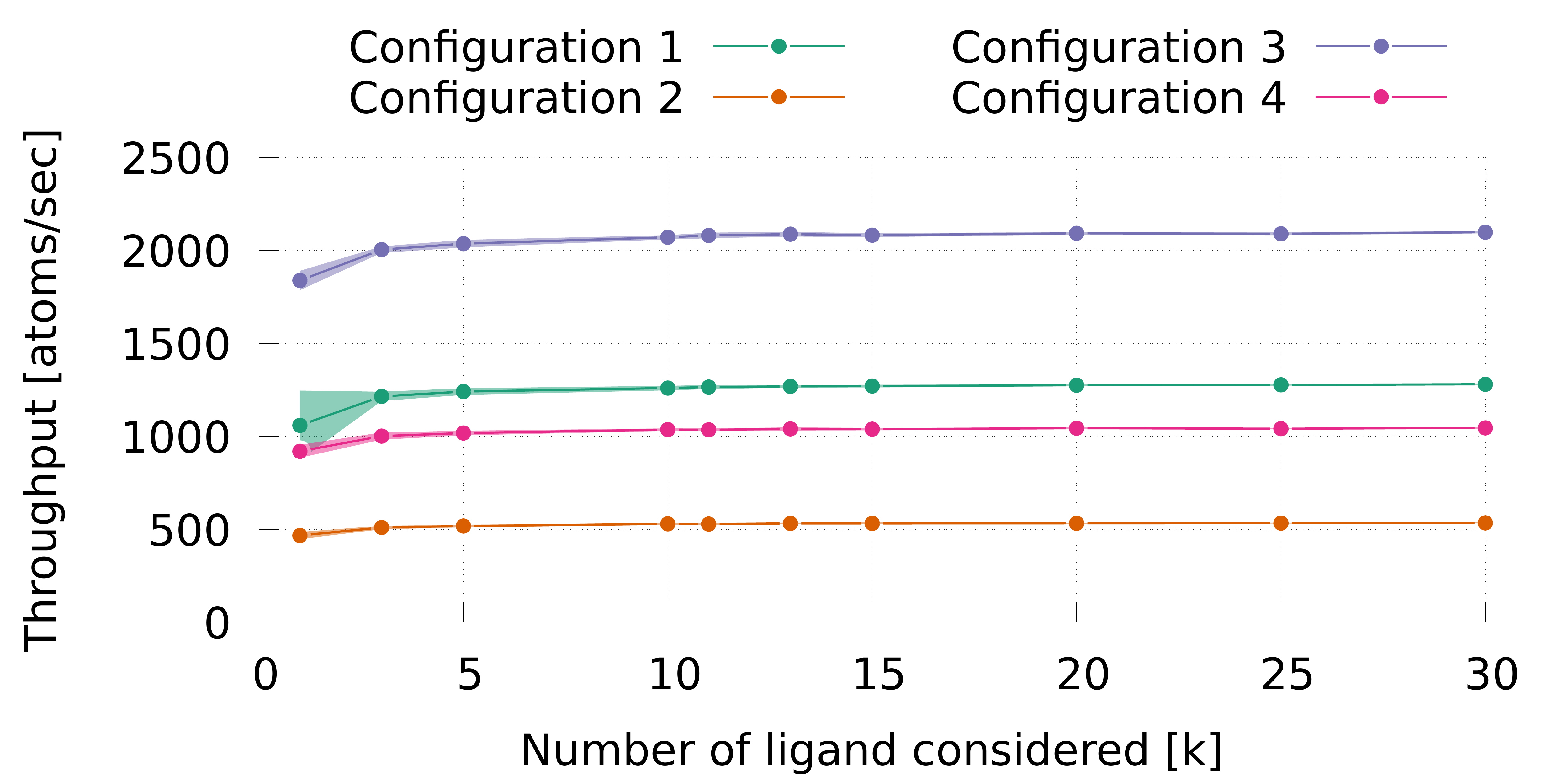}}\\
	\centering
	\subfloat[Overlap score degradation]{%
		\label{fig:galileo_error_plot}%
        \includegraphics[width=1.0\columnwidth]{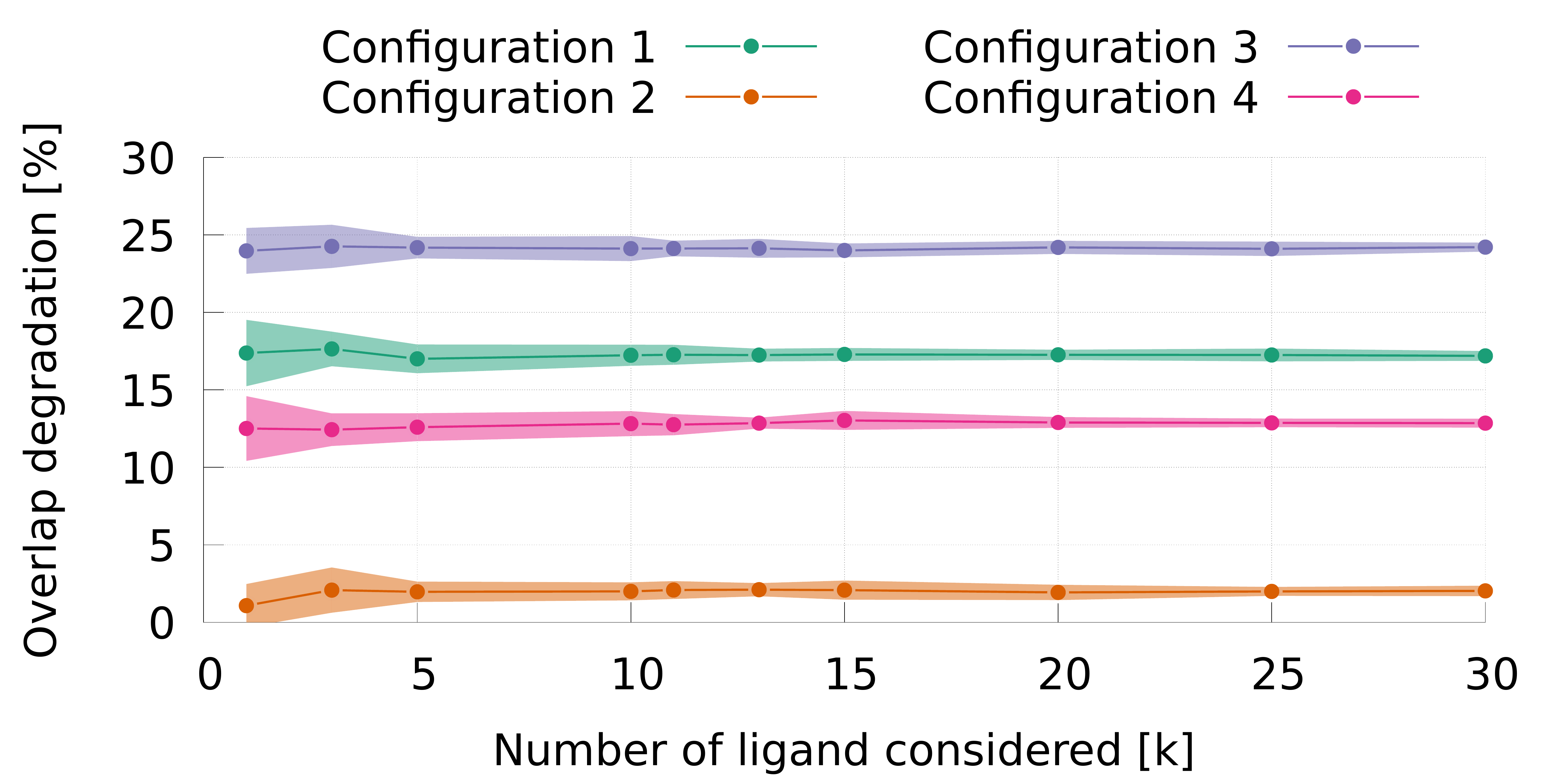}}\\
\caption{Application analysis in terms of throughput per process and overlap score degradation by varying the number of ligands. For each configuration we show the average values (dot) and the standard deviation (colored area).}
\label{fig:approx-approach}
\end{figure}

In this paper, we enhanced the geometrical docking module of LiGenDock with approximation techniques, to trade off the quality of the results with respect to the throughput.
Therefore we are interested in finding the set of configurations in the Pareto front, given by the non-dominated solutions considering both target metrics. 
However, the application needs to find the most promising ligands across a heterogeneous data set.
This means that the performance of the application might depend on which ligands are considered.

To avoid the profiling phase of the alternative configurations for each database of ligands that we want to evaluate, this experiment wants to assess how much the performance of the application is dependent on the changes in the dataset.
To this end, we evaluate four different configurations of the enhanced version of \MiniApp{} in terms of tunable knobs.
For each configuration, we characterize the application behaviour in terms of throughput and overlap degradation by varying the number of considered ligands.
The set of ligands considered to evaluate each configuration has been selected by randomizing 20 times over the full set of 113K elements, thus emulating new datasets.

\prettyref{fig:approx-approach} shows the results of the experiment.
In particular, \prettyref{fig:galileo_throughput_plot} focuses on the application throughput (y-axis), while \prettyref{fig:galileo_error_plot} focuses on the overlap degradation (y-axis).
For both of them, each dot represents the mean performance of the evaluated configuration by varying different databases of ligands.
The transparent solid area represents the uncertainty of the mean by using the standard deviation of the measures.
The x-axis indicates the number of ligands considered in the evaluation.

From these results we see in \prettyref{fig:galileo_throughput_plot} that the average application throughput has a very limited dependency on both the number of ligands in the target database and the input data (i.e. very small standard deviation).
This is an expected result, because the throughput definition we considered is related to the number of atoms of the database instead of the number of ligands, thus providing a normalized measure.
On the other hand, \prettyref{fig:galileo_error_plot} shows how the overlap degradation is a little bit more data dependent than the throughput.
In particular, we need to consider at least $5K$ ligands to have a steady average value.
This behaviour is due to the overlap degradation definition that makes the value dependent on the top $1\%$ ligands and therefore on which ligands are selected. 
However, we can determine that less than 10K ligands are enough to characterize the configurations of the enhanced version of \MiniApp{} for both throughput and overlap degradation.

\subsection{Trade-off Analysis}
\label{subsec:dse}

\begin{figure}
	\centering
		
        \includegraphics[width=1.0\columnwidth]{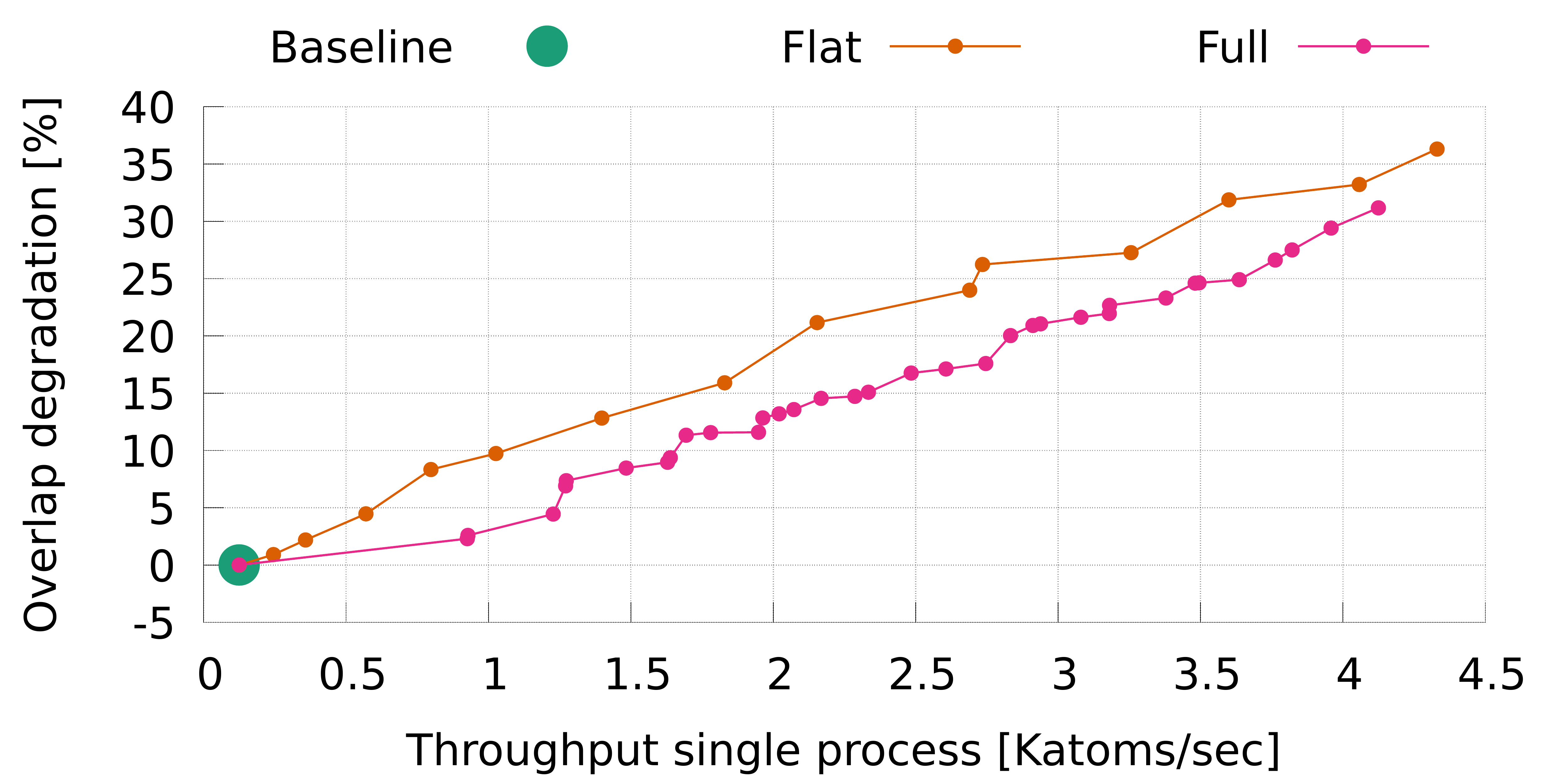}
\caption{Pareto front of \MiniApp{} in terms of overlap score degradation versus throughput: \emph{Flat} and \emph{Full} \label{fig:dse_final}}
\label{fig:approx-dse}
\end{figure}

This experiment wants to show performance-accuracy tradeoffs when we apply the approximation techniques described in \prettyref{subsec:approx}.
\prettyref{fig:approx-dse} shows the Pareto front of the design space exploration carried out in a single node of Galileo using 20k ligands, targeting a single pocket.

\prettyref{fig:dse_final} compares the performance of the application by using a \emph{flat} sampling on the rotation angles, as proposed in the LiGenDock paper \cite{beato2013use}, with the performance of the application using the \emph{full} set of software knobs proposed in this paper.
In particular, the \textit{flat} design space is given by:
\param{high-precision step} $[1^\circ, 2^\circ, 3^\circ, 5^\circ, 10^\circ, 15^\circ, 45^\circ, 60^\circ]$,
\param{repetitions} [1, 2, 3].
The \textit{full} design space (for evaluating all the software knobs proposed in the paper) is given by:
\param{high-precision step} $[1^\circ,2^\circ,3^\circ,5^\circ]$,
\param{low-precision step} $[45^\circ,90^\circ]$,
\param{threshold} [0, 0.3, 0.6, 0.8],
\param{repetitions} [1, 2, 3],
\param{enable refinement} [$true$, $false$].
For both design spaces, we used a full factorial Design of Experiments. 
We considered as \emph{Baseline} configuration the most precise one, that is exactly the same version that can be derived either from \emph{flat} or \emph{full} -- i.e. \param{high-precision step}=$1^\circ$ and \param{repetitions}=3 for the \emph{flat} version, and \param{high-precision step}=$1^\circ$, \param{low-precision step}=*, \param{threshold}=0, \param{repetitions}=3, and \param{enable refinement}=$false$ for the \emph{full} version.

As expected, from the results reported in \prettyref{fig:dse_final}, the Pareto front derived by the \emph{full} version dominates the one derived by the \emph{flat} sampling.
It is possible to notice how only by enabling the iterative refinement (first configuration on the \emph{full} curve after the \emph{Baseline}), we can greatly improve the throughput of the application (7.4X) with a limited overlap degradation (2.3\%) with respect to the \emph{Baseline} version. 

\begin{figure}
	\centering
    \includegraphics[width=0.99\columnwidth]{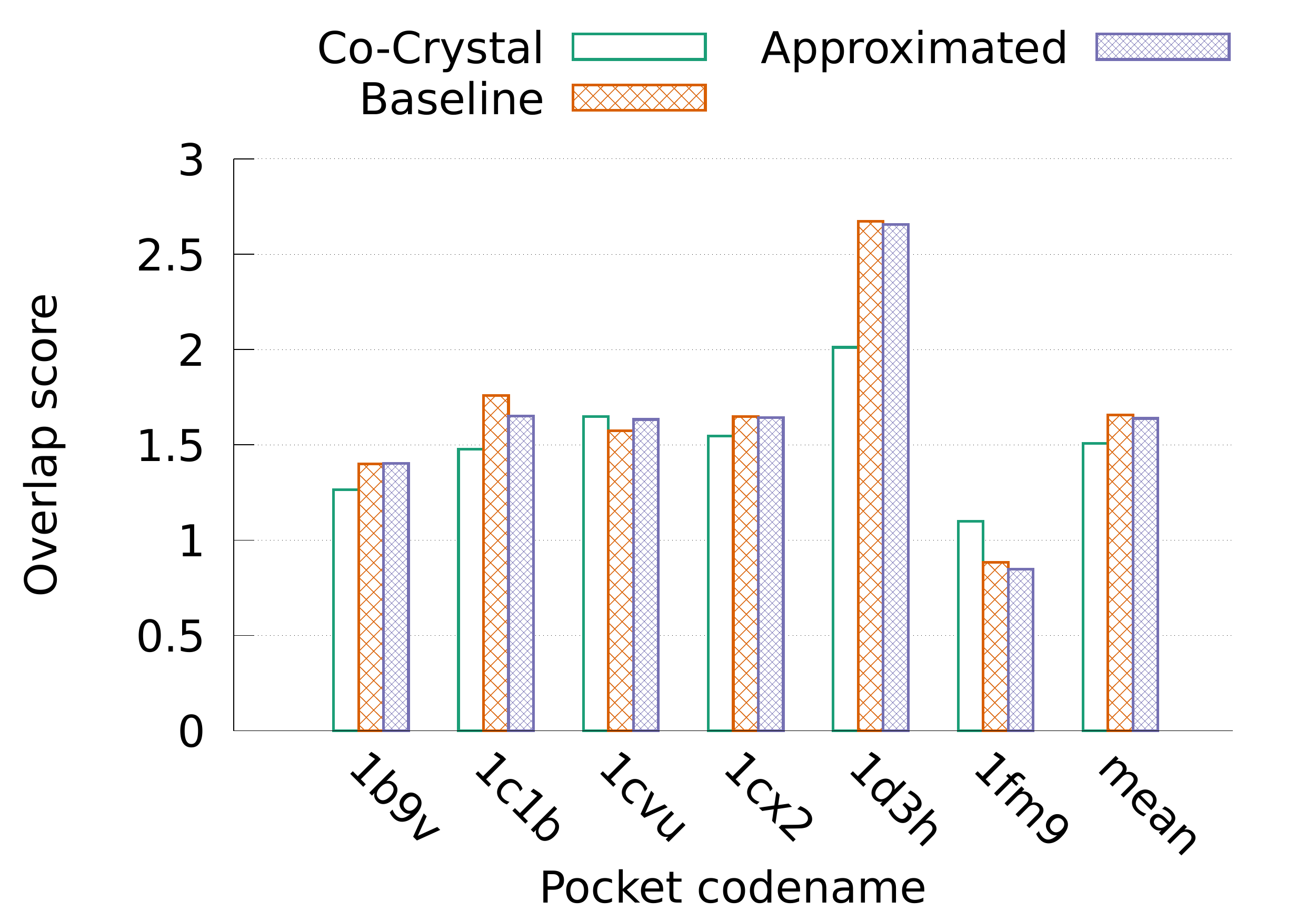}
\caption{Overlap score by varying the target pocket for the co-crystallized ligand, for the baseline and proposed approximated version. \label{fig:comparison}}
\end{figure}

The Protein Data Bank (PDB) \cite{Berman00theprotein} contains the three-dimensional structural data of biological molecules, providing also the pose of the ligand when co-crystallized within the target pocket (i.e. the actual pose of the best ligand for that pocket). 
Therefore, we decided to use some pocket-ligand pairs to better analyze the effects of accuracy degradation.
\prettyref{fig:comparison} shows for each pocket-ligand pair: a) the overlap score of the crystal, b) the overlap score of the docked ligand using the \emph{baseline} version, and c) the overlap score of the approximated version using only the iterative refinement (i.e. 7.4x speedup).
We may notice that the investigated configuration of the enhanced \MiniApp{} has a small degradation of the overlap score not only in the average case but also when considering a single target ligand.
The overlap score of the co-crystallized ligand is usually lower with respect to the computed ones because the real pose of the ligand takes into consideration also chemical features which are not captured by the geometrical score.

\subsection{Time-to-solution Model Validation}
This experiment aims at validating the time-to-solution model described in \prettyref{subsec:application_autotuning}.
In particular, the model is defined in the design space of the \textit{full} version described in \prettyref{subsec:dse}.
To compute the coefficients of the linear regression for each configuration (see \prettyref{eq:linear_regression}), we run the application several times by using $1K$ ligands per pocket for each configuration.
The extracted models for each configuration have an average adjusted $R^2$ value equal to 0.977.

\begin{figure}[t!]
	\centering
    \includegraphics[width=1.0\columnwidth]{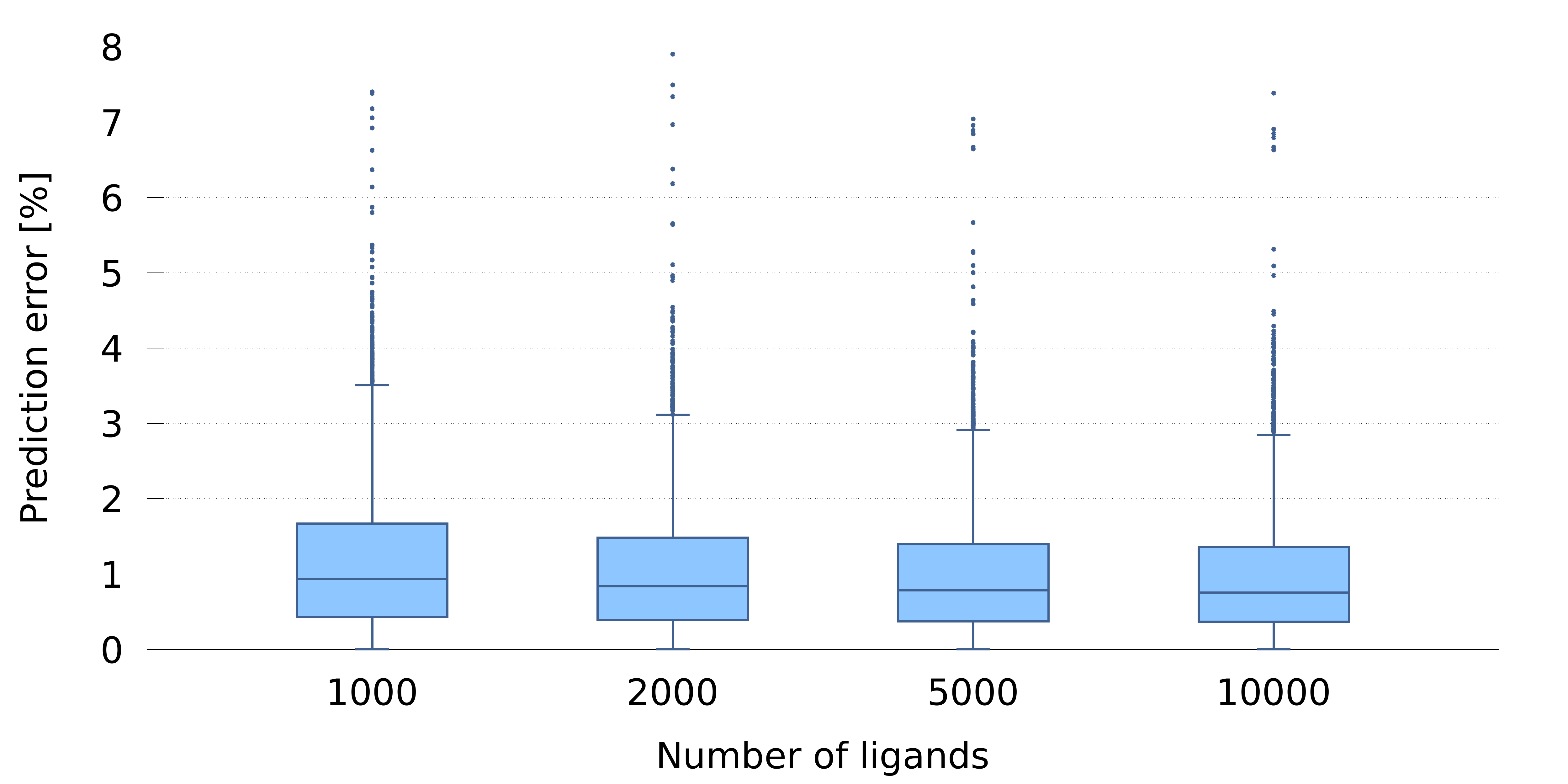}
\caption{Prediction error of the time-to-solution, by varying the number of ligands to dock in the target pocket.\label{fig:model_validation}}
\end{figure}

We run an experimental campaign to further validate the accuracy of the time-to-solution prediction of the model, by using the leave-one-out scheme on the pockets and a different set of ligands with respect to those used for the training.
For each pocket and configuration stated in \prettyref{subsec:dse}, we executed the application with three different databases randomly selected over the entire set and composed of $1k$, $2k$, $5k$ and $10k$ ligands. We did not use very small sizes for the target database because our goal is to predict the time-to-solution during a virtual-screening process, typically composed of a large number of target ligands.
For each run, we stored the predictor value of the input and the observed time-to-solution to extract the prediction error.
\prettyref{fig:model_validation} shows the distribution of the prediction error of our model by varying the number of ligands in the experiment.
The average error is below $1\%$ of the observed time-to-solution for the entire range of considered ligand-database size, while the outliers (we validate the application by using more than $15K$ experiments) reache a maximum value of $7.9\%$, with a trend which is stable by increasing the number of docked ligands.

\subsection{Use-case Scenarios}
\label{subsec:use_case_scenario}

\begin{figure*}[t!]
	\centering
	\subfloat[Scenario 1: Varying the size of the ligand database given one-day time budget.]{
		\label{fig:ttos_size}
        \includegraphics[width=0.95\columnwidth]{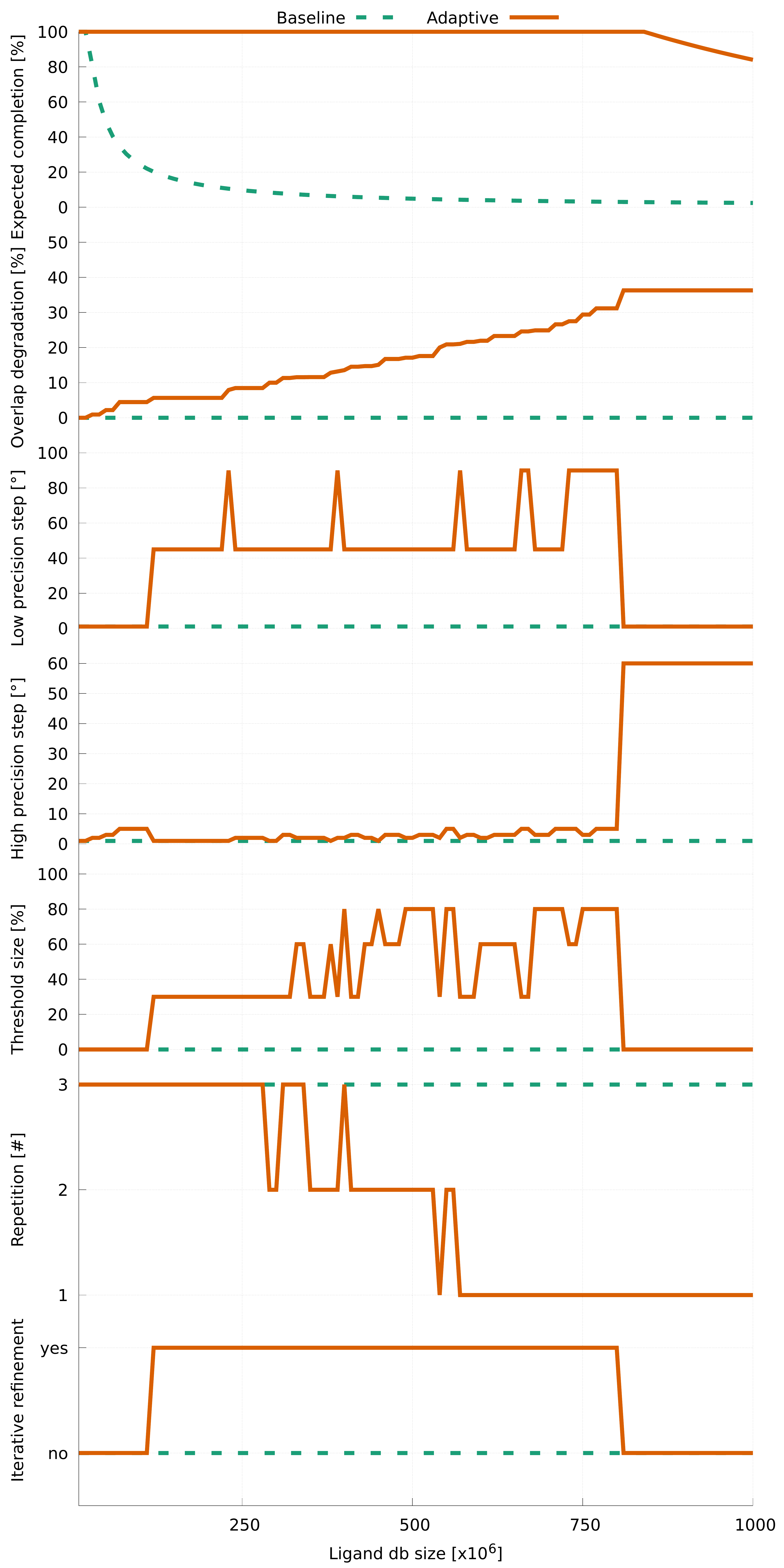}}
        \hspace{0.8cm}
    \subfloat[Scenario 2: Varying the allocated time budget given the ligand database size equal to $500\times10^6$.]{
		\label{fig:ttos_ttos}
        \includegraphics[width=0.95\columnwidth]{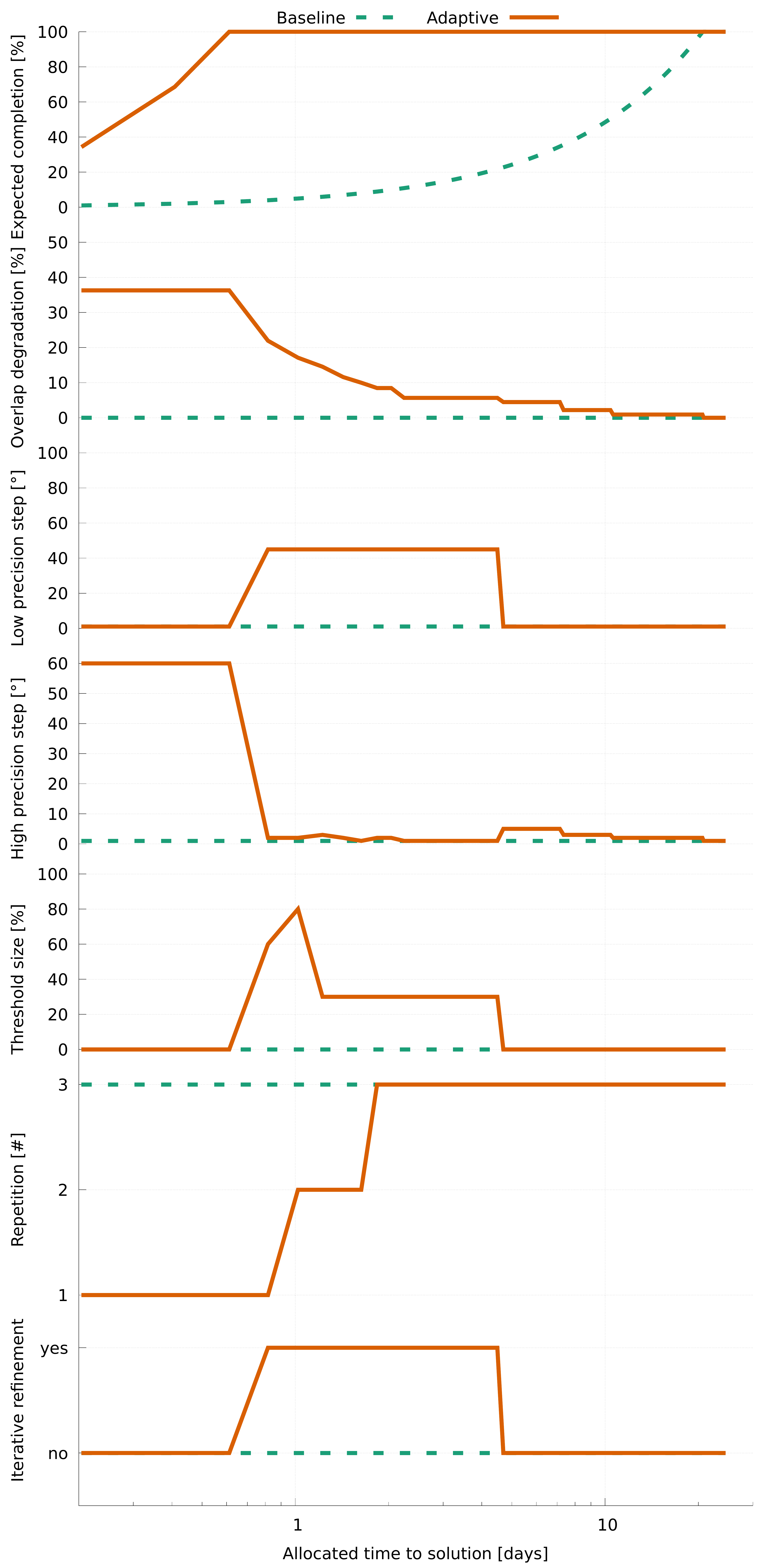}}
\caption{\MiniApp{} behavior in terms of expected percentage of ligand database completion, expected overlap degradation, and the selected configuration (a) by varying the size of the input, and (b) the time budget, when using 8 nodes of Galileo supercomputer.}
\label{fig:ttos}
\end{figure*}

The last experiment wants to evaluate the benefits of the proposed approach for the end-user, i.e. a pharmaceutical company that aims at screening a large set of ligands given a time budget. 
We envisioned two use cases to exploit the performance-accuracy tradeoff.
In the first scenario, we considered a fixed time budget for the computation and we would like to understand what is the effect of incrementing the size of the database to be analyzed, in order to increase the chances to find a better drug.
In the second scenario, we fixed the size of the database and we observed the effects of varying the time budget, thus varying the cost of the experiment.
We can summarize these two scenarios as attempts to offer to the end-user with two high-level knobs: In the first case the number of ligands to be screened, while in the second case the time budget (i.e. the cost of the experiment). The time-to-solution model will be used to set the right low-level application knobs included in \MiniApp{} while satisfying the constraints. 

\prettyref{fig:ttos} shows the expected behaviour of the application, assuming that the end-user is using 8 nodes of the Galileo supercomputer (see \prettyref{subsec:target}).
On the y-axis are represented the expected performance of the application (top 2 plots) and the selected configuration of the software knobs (bottom 5 plots).
The performance of the application is defined in terms of the expected completion percentage of the planned ligand database and the related overlap degradation of the result. The completion percentage is the ratio between the number of ligands docked given the time budget and the size of the target ligand database.
Each plot includes two lines: one dotted line for the baseline version of \MiniApp{} and the other line for the adaptive version proposed in this paper.

The x-axis represents the high-level parameter tuned by the end-user, according to the scenario.
In Scenario 1 (\prettyref{fig:ttos_size}), the end-user would like to tune the size of the database of ligands to complete the job given one-day time budget.
In Scenario 2 (\prettyref{fig:ttos_ttos}), the end-user would like to tune the time budget of the application to dock a given database of $500\times10^6$ ligands.

In both scenarios, we notice that the proposed software knobs enable a swing of several orders of magnitude for the end-user to tune the parameters of the job (i.e. problem size and time-to-solution).
Moreover, by using an autotuner together with the time-to-solution model, we are capable to alleviate the burden of the manual selection of the software knobs from the end-user, by exposing simpler parameters. Although the average trend of the application knobs values for both scenarios can be derived from their meaning, the actual values and when to switch among the configurations according to the high-level constraints (i.e. problem size and time-to-solution) is hard to know without any automatic support.
A clear example of this is the behaviour of the \param{threshold}-value within the middle range (300--700$\times10^6$) of the problem size considered in the experiment shown by \prettyref{fig:ttos_size}.

In terms of application performance, we can notice how in the first scenario, where we kept constant the time to solution (see \prettyref{fig:ttos_size}), the proposed \emph{adaptive} version is able to process the whole database up to 850$\times10^6$ ligands, the baseline version rapidly decreases the rate to very low values when enlarging the database size.
This happens thanks to the capability of the proposed \emph{adaptive} version, not present in the \emph{baseline}, to trade-off accuracy and performance. Moreover, the second plot in \prettyref{fig:ttos_size} also demonstrates how the deep parametrization introduced in the target Mini-App provides a smooth degradation of the application quality.
In the second scenario shown in \prettyref{fig:ttos_ttos}, where we fixed the size of the target database and we varied the allocated time budget, we observed a similar behaviour. In fact, increasing the time budget, the proposed \emph{adaptive} version rapidly reaches ($<1$ day) the value of 100\% completion at the cost of a low accuracy, while for the baseline we had to allocate more than 20 days.

Finally, in both scenarios we can identify when the ligand database size, or the allocated budget, is set to values that guarantee only $10\%$ of the expected completion for the \emph{baseline} version, the \emph{adaptive} one results with a limited degradation less than $10\%$. This demonstrates the effectiveness of the extracted low-level knobs in \MiniApp{}.

\section{Related Work}
\label{sec:relate}
Being our approach intrinsically interdisciplinary, we partitioned the relevant literature in three sections covering the research areas related to this work: molecular docking, approximation techniques and application autotuning.

\textbf{Molecular docking techniques.} 
Molecular docking is a well-known research topic addressed in literature from different perspectives.
A large set of works addresses the problem by exploiting random-based algorithms, such as genetic algorithms \cite{thomsen2006moldock,jones1997development} or Monte Carlo simulations \cite{friesner2004glide,liu1999mcdock}.
However, a desirable feature of a molecular docking application is the determinism of the solution.
Because of the tasks following the \emph{in-silico} step require expensive solution-dependent resources, for several companies having a deterministic and repeatable result is a constraint.
Early works in literature, such as \cite{jiang1991soft}, generate deterministic solutions.
However, they consider only rigid movements of the ligand during the docking procedure.
Real case scenarios usually require the rotation of portions of the ligand molecule.
Therefore, the limitation of rigid movements is likely to prevent the applicability of the solution in the industry.
The work of Palma et al.~\cite{palma2000bigger} overcomes this issue: they introduced a molecular docking framework to deal with the flexibility of the ligand molecule.
It adopts a model of the electrostatic interactions to finalize the docking.
Similar works such as DOCK~\cite{ewing2001dock}, FlexX~\cite{Kramer1999flexx}, FlexX-Scan~\cite{schellhammer2004flexx} and sur-flex~\cite{jain2007surflex} provide deterministic molecular docking of flexible ligands.
They enabled the user to exploit several docking algorithms according to the specific use case.
All these algorithms rely on both geometric and pharmacophoric properties in their docking algorithms.
All these works implement a different docking procedure with respect to LiGenDock. However, no one of them has been designed to expose software knobs to enable explicitly  quality-performance tradeoffs. 
The proposed approach behind \MiniApp{} unveil the possibility to tune the docking procedure according to high-level constraints, such as the allocated time budget for a given size of the ligand database to be virtually screened.

\textbf{Approximation techniques.} During the last years, the most promising approaches to achieve performance gains are represented by the trade-off with application accuracy~\cite{8342176}, \cite{Esmaeilzadeh:2012:ASD:2150976.2151008}.
Algorithm-level approximate computing techniques are well-known in literature~\cite{6569370}, \cite{XU16AxSurvey}, \cite{Mittal16AxSurvey}, and represent also an important challenge for in HPC applications~\cite{6877311}, \cite{Vassiliadis2016ExploitingSO}.
In this work, we exploit grid-based optimizations on the geometrical docking kernel.
In particular, in computational physics, it is very common to exploit models based on multi-level grids to achieve a fine-grained solution in a restricted area of the whole simulated environment.
The size of the grid is a parameter which enables the physicists to trade-off granularity of solution for the increasing/decreasing number of elements to be processed.
Nested grids have been exploited for a long time in geophysics applications, such as thermosphere models~\cite{fuller1984two,fuller1985two,wang1999high} and ocean flows models~\cite{ocean1992}.
The evolution of nested-grids models made the researchers to abandon regular grids in favour of variable-size grids.
Irregular grids have been exploited in climate forecast models to improve the performance of grid-based models.
Authors of~\cite{climate2001} demonstrate that a variable-resolution stretched grids lead to longer-term climate forecast with the same accuracy of the nested grid models.
In physics, variable-size grids are used to discretize geophysical problems such as advection equations~\cite{ullrich2011analysis}.
In the field of image rendering, grid processing has been optimized by selecting which tiles need to be processed first and which one later or do not require any processing at all. An element is peeled from each tile, and its value is used to decide how to compute the corresponding tile, e.g. deep peeling~\cite{graphic_raycast2006, everitt2001interactive,earthquake3D}.

\textbf{Application autotuning.}
Autotuning in HPC is a well-known research area \cite{AutotuningInHPC} where several libraries and tools exist to support the automatic optimization of software knobs.
Among the libraries that are tailored to specific tasks some examples are ATLAS \cite{whaley1998automatically} for matrix multiplication routines, FTTW \cite{frigo2005design} for FFT operations, OSKI \cite{vuduc2005oski} for sparse matrix kernels, and SPIRAL \cite{puschel2004spiral} for digital signal processing.
Frameworks such as \cite{ansel2011language,ORIO,PTF,ACTIVEARMONY} consider the autotuning problem from a more general perspective. They have been designed to make an application tunable without being bound to a target application field. 
However, only a few approaches exist to deal with approximation knobs \cite{DYNAMIC_KNOBS, CAPRI}, end even fewer frameworks consider data-aware approximation \cite{PETABRICKS_input,Samadi:2013:SSA:2540708.2540711,GRAPHTUNER}. 
The work in \cite{PETABRICKS_input} enhances Petabricks \cite{ansel2011language} to leverage the accuracy-throughput tradeoffs and to consider input features of the data to be processed, On the other side, \cite{Samadi:2013:SSA:2540708.2540711} and \cite{GRAPHTUNER} follows the same idea of switching at runtime the configuration according to the input characteristics by applying it to two different application domains, respectively graphics and graph processing.
%
%
%
To support the autotuning phase, machine learning techniques have been used in literature \cite{CAPRI, MODEL1, MODEL2} to model application metrics and then to predict the best configuration to be applied. This is fundamental when the size of the configuration space is huge, and thus not possible to be entirely profiled, or when the elaboration is heavily data-dependent. The workload we are considering in this paper falls in this second case and these previous works have been used as inspiration to our predictive solution.
%


\section{Conclusion}
\label{sec:conclusion}

Virtual screening is a crucial task of a drug discovery process as it aims at finding the most promising molecules to evaluate on the later stage of the process.
Due to the large number of theoretical molecules that may be evaluated, any speedup of the task leads to either a reduction of the economic cost of the whole process or the possibility to evaluate a higher number of possible candidates.
In this paper, we have analyzed \MiniApp{}, which represents the most computational heavy kernels of LiGenDock when it is used to perform a screening on a target database of ligands. 
From the analysis of \MiniApp{}, we identified several approximation possibilities, declined in five software knobs, enabling the computation to gain accuracy only when it is likely to have more impact on the output, thus enabling accuracy-performance tradeoffs.
The adaptive version of \MiniApp{} provides different accuracy levels according to the evolving needs of the virtual screening experimental campaign.
In particular, experimental results demonstrated how, by playing with the introduced software knobs, the proposed tunable application adaptively runs the virtual screening campaign over a ligand database by introducing an accuracy degradation of less than $10\%$ during the same time used by the original version to screen only $1/10$ of the database.
Due to the determinism of the proposed algorithm and the software knobs, we defined a prediction model to support the application autotuning. The model is capable to autotune low-level knobs of the application by exposing to the end-user the high-level parameters (such as the time budget, the size of the ligand database and the number of resources).
These outcomes represent a huge advantage for pharmaceutical industries in a context where the use of HPC systems and software in drug discovery have become crucial assets to become more and more competitive in the drug discovery process.

\ifCLASSOPTIONcaptionsoff
  \newpage
\fi

\bibliographystyle{unsrt}
\bibliography{references}

\end{document}